\documentclass[aps,prl,twocolumn,showpacs,superscriptaddress,twoside,showpacs]{revtex4-1}
\usepackage{CJK}
\usepackage{graphicx}
\usepackage{dcolumn}
\usepackage{longtable}
\usepackage{bm}
\usepackage{booktabs}
\usepackage{amsmath}
\usepackage{color}

\begin{document}
\title{Diversity of shapes and rotations in the $\gamma$-soft $^{130}$Ba nucleus: first observation of a $t$-band in the A=130 mass region}

\author{C. M. Petrache}
\thanks{{\it Corresponding author}}
\email{petrache@csnsm.in2p3.fr}
\affiliation{Centre de Sciences Nucl\'eaires et Sciences de la
 Mati\`ere, CNRS/IN2P3, Universit\'{e} Paris-Saclay, B\^at. 104-108, 91405  Orsay, France}
 \author{P. M. Walker}
\affiliation{Department of Physics, University of Surrey, Guildford GU2 7XH, United Kingdom}
 \author{S. Guo}
\affiliation{Institute of Modern Physics, Chinese Academy of Sciences, Lanzhou 730000, China}
\author{Q. B. Chen}
\affiliation{Physik-Department, Technische Universit\"{a}t M\"{u}nchen, D-85747 Garching, Germany}
\author{S. Frauendorf}
\affiliation{Department of Physics, University Notre Dame, Indiana 46557, USA} 
 \author{Y. X. Liu}
\affiliation{Department of Physics, Huzhou University, Huzhou 313000, China}
\author{R. A. Wyss}
\affiliation{KTH Department of Physics,S-10691 Stockholm, Sweden}
\author{D. Mengoni}
\affiliation{Dipartimento di Fisica e Astronomia dell'Universit\`a, and INFN, Sezione di Padova, 35131 Padova, Italy}
 \author{Y. Qiang}
\affiliation{Institute of Modern Physics, Chinese Academy of Sciences, Lanzhou 730000, China}
\author{A. Astier}
\affiliation{Centre de Sciences Nucl\'eaires et Sciences de la
  Mati\`ere, CNRS/IN2P3, Universit\'{e} Paris-Saclay, B\^at. 104-108, 91405  Orsay, France}
\author{E. Dupont}
\affiliation{Centre de Sciences Nucl\'eaires et Sciences de la
  Mati\`ere, CNRS/IN2P3, Universit\'{e} Paris-Saclay, B\^at. 104-108, 91405  Orsay, France}
  \author{R. Li}
\affiliation{Centre de Sciences Nucl\'eaires et Sciences de la
  Mati\`ere, CNRS/IN2P3, Universit\'{e} Paris-Saclay, B\^at. 104-108, 91405  Orsay, France}
    \author{B. F. Lv}
\affiliation{Centre de Sciences Nucl\'eaires et Sciences de la
  Mati\`ere, CNRS/IN2P3, Universit\'{e} Paris-Saclay, B\^at. 104-108, 91405  Orsay, France}
  \author{K. K. Zheng}
\affiliation{Centre de Sciences Nucl\'eaires et Sciences de la
  Mati\`ere, CNRS/IN2P3, Universit\'{e} Paris-Saclay, B\^at. 104-108, 91405  Orsay, France}
 \author{D. Bazzacco}
\affiliation{Dipartimento di Fisica e Astronomia dell'Universit\`a, and INFN, Sezione di Padova, 35131 Padova, Italy}
\author{A. Boso}
\affiliation{Dipartimento di Fisica e Astronomia dell'Universit\`a, and INFN, Sezione di Padova, 35131 Padova, Italy}
\author{A. Goasduff}
\affiliation{Dipartimento di Fisica e Astronomia dell'Universit\`a, and INFN, Sezione di Padova, 35131 Padova, Italy}
\author{F. Recchia}
\affiliation{Dipartimento di Fisica e Astronomia dell'Universit\`a, and INFN, Sezione di Padova, 35131 Padova, Italy}
\author{D. Testov}
\affiliation{Dipartimento di Fisica e Astronomia dell'Universit\`a, and INFN, Sezione di Padova, 35131 Padova, Italy}
\author{F. Galtarossa}
\affiliation{INFN Laboratori Nazionali di Legnaro, 35020 Legnaro (Pd), Italy}
\author{G. Jaworski}
\affiliation{INFN Laboratori Nazionali di Legnaro, 35020 Legnaro (Pd), Italy}
\author{D. R. Napoli}
\affiliation{INFN Laboratori Nazionali di Legnaro, 35020 Legnaro (Pd), Italy}
\author{S. Riccetto}
\affiliation{INFN Laboratori Nazionali di Legnaro, 35020 Legnaro (Pd), Italy}
\author{M. Siciliano}
\affiliation{INFN Laboratori Nazionali di Legnaro, 35020 Legnaro (Pd), Italy}
\affiliation{IRFU, CEA, Université Paris-Saclay, 91191 Gif-sur-Yvette, France}
\author{J. J. Valiente-Dobon}
\affiliation{INFN Laboratori Nazionali di Legnaro, 35020 Legnaro (Pd), Italy}
\author{M. L. Liu}
\affiliation{Institute of Modern Physics, Chinese Academy of Sciences, Lanzhou 730000, China}
 \author{X. H. Zhou}
\affiliation{Institute of Modern Physics, Chinese Academy of Sciences, Lanzhou 730000, China}
 \author{J. G. Wang}
\affiliation{Institute of Modern Physics, Chinese Academy of Sciences, Lanzhou 730000, China}
 \author{C. Andreoiu}
\affiliation{Department of Chemistry, Simon Fraser University, Burnaby, BC V5A 1S6, Canada} 
\author{F. H. Garcia}
\affiliation{Department of Chemistry, Simon Fraser University, Burnaby, BC V5A 1S6, Canada}
\author{K. Ortner}
\affiliation{Department of Chemistry, Simon Fraser University, Burnaby, BC V5A 1S6, Canada}
\author{K. Whitmore}
\affiliation{Department of Chemistry, Simon Fraser University, Burnaby, BC V5A 1S6, Canada}
\author{T. B\"ack}
\affiliation{KTH Department of Physics,S-10691 Stockholm, Sweden}
  \author{B. Cederwall}
\affiliation{KTH Department of Physics,S-10691 Stockholm, Sweden}
\author{E. A. Lawrie}
\affiliation{iThemba LABS, National Research Foundation, PO Box 722, 7131 Somerset West, South Africa}
\author{I. Kuti}
\affiliation{Institute for Nuclear Research, Hungarian Academy of Sciences, Pf. 51, 4001 Debrecen, Hungary}
\author{D. Sohler}
\affiliation{Institute for Nuclear Research, Hungarian Academy of Sciences, Pf. 51, 4001 Debrecen, Hungary}
\author{J. Tim\'ar}
\affiliation{Institute for Nuclear Research, Hungarian Academy of Sciences, Pf. 51, 4001 Debrecen, Hungary}
\author{T. Marchlewski}
\affiliation{Heavy Ion Laboratory, University of Warsaw, 02-093 Warsaw, Poland}
\author{J. Srebrny}
\affiliation{Heavy Ion Laboratory, University of Warsaw, 02-093 Warsaw, Poland}
\author{A. Tucholski} 
\affiliation{Heavy Ion Laboratory, University of Warsaw, 02-093 Warsaw, Poland}

\begin{abstract}

Several new bands have been identified in  $^{130}$Ba, among which there is one with band-head spin $8^+$. Its properties are in agreement with the Fermi-aligned $\nu h_{11/2}^{2}$, $7/2^- [523] \otimes 9/2^-[514]$ Nilsson configuration. This is the first observation of a two-quasiparticle $t$-band in the A=130 mass region. The $t$-band is fed by a dipole band involving two additional $h_{11/2}$ protons. The odd-spin partners of the proton and neutron $S$-bands and the ground-state band at high spins are also newly identified. The observed bands are discussed using several theoretical models, which strongly suggest the coexistence of prolate and oblate shapes polarized by rotation aligned two-proton and two-neutron configurations, as well as prolate collective rotations around axes with different orientations. With the new results, $^{130}$Ba presents one of the best and most complete sets of collective excitations that a $\gamma$-soft nucleus can manifest at medium and high spins, revealing a diversity of shapes and rotations for the nuclei in the $A=130$ mass region.
 
  \end{abstract}



\maketitle

\section{Introduction}{\label{int}}

The nuclei of the $A=130$ mass region around $N=76$ often have properties in agreement with a $\gamma$-soft triaxial shape at low spins, while at medium and high spins the shape can change to nearly axially symmetric as for the high-$K$ isomers \cite{130ba-isomer}, and highly-deformed or superdeformed bands (see e. g. \cite{trs1988,138nd-low}). States with spins higher than $10^+$ in even-even nuclei are built by breaking one nucleon pair with alignment of the spins of the particles or holes along the rotation axis. 
The spins of the low-$\Omega$ protons at the prolate (or nearly prolate) Fermi surface are rapidly aligned parallel to the rotation axis under the influence of the Coriolis force, giving rise to {\it rotation-aligned} (RAL) bands called $S$-bands, while the corresponding high-$\Omega$ neutrons ({\it i. e.} for prolate shape) are strongly coupled and their spins remain parallel to the long axis, giving rise to {\it deformation-aligned} (DAL) rotational bands called $K$-bands. However, the shape of $\gamma$-soft nuclei can become oblate under the polarizing effect of the neutron holes, and in such a case one can also obtain an $S$-band by rotation alignment of the neutron holes.
 As the structure and orientation axis in the two-neutron high-$K$ configurations
are very different from those of the ground-state low-$K$ band, the band-heads have often long lifetimes and can
become isomeric. This situation has been recently discussed in $^{130}$Ba \cite{130ba-isomer}, in which a dipole band built on the long-lived $K^{\pi}= 8^-$ isomer has been observed, having properties in agreement  with nearly axially symmetric prolate shape. Such high-$K$ isomers are also known in the deformed rare-earth nuclei with A $\approx160 - 180$, like the Os and W nuclei \cite{Dr16}. The $^{130}$Ba nucleus also exhibits a $\gamma$-band with strong energy staggering between the odd and even spins, a fingerprint of the $\gamma$-softness or $O(6)$ symmetry (see  \cite{sevrin1987,kirch,caprio}). This softness facilitates the coexistence of different shapes, which can change depending on the specific configuration, ranging from prolate to oblate. In addition to the RAL  $S$-bands and DAL
$K$-bands, there can exist a special type of bands, called $t$-bands, intermediate between the DAL and RAL regimes, in which the cranking axis is tilted away from the principal axes of the spheroidal core. Such bands originate  from high-$j$ quasiparticles with the 
Fermi level in the middle of the shells, which were called {\it Fermi-aligned} (FAL) by Frauendorf \cite{fra_tilted-cranking,fra-2000}. 
 One example is $^{180}$W, in which a rotational band built on the $8^+$, 2132-keV state has
been interpreted as a $\nu i^2_{13/2}$  FAL $t$-band \cite{180W-walker}, and another is $^{182}$Os, in which two bands built on $8^+$ and $9^+$ states are also interpreted as a $\nu i^2_{13/2}$ FAL $t$-band \cite{182os-1995}. No two-quasiparticle $t$-bands were reported until now in the $A\approx130$ mass region, even though the Fermi level for $N\approx74$ can be mid-way among the $h_{11/2}$ orbitals and favor $t$-band configurations.  

In the present letter we report for the first time in the $A\approx130$ mass region the observation in $^{130}$Ba of a band with characteristics similar to those of the FAL configurations assigned to the $t$-bands. New experimental information is also reported on several medium-spin bands related to the $t$-band, in particular odd-spin partners of the $S$-bands and the continuation at high spins of the ground state band. Several theoretical models have been used to interpret the observed structures: Total Routhian Surface (TRS) \cite{trs1988}, tilted axis cranking (TAC) \cite{fra_tilted-cranking}, particle rotor model (PRM) \cite{Ch18}, and projected shell model (PSM) \cite{hara-sun,Sun-psm}.
 
\section{Experimental results}{\label{exp}}

The $^{130}$Ba nucleus has been populated via the $^{122}$Sn($^{13}$C,5n) reaction at a beam energy of 65 MeV. The
target consisted of a stack of two self-supporting $^{122}$Sn foils with a thickness of 0.5 mg/cm$^2$ each. The $^{13}$C beam
of 5 pnA was provided by the XTU Tandem accelerator of the Laboratori Nazionali di Legnaro. The $\gamma$-rays were detected
by the GALILEO spectrometer \cite{val2014,testov2018}, which consisted of 25 Compton-suppressed Ge detectors placed on four rings at 90$^{\circ}$ (10 detectors), 119$^{\circ}$ (5 detectors), 129$^{\circ}$ (5 detectors) and 152$^{\circ}$ (5 detectors). To distinguish different reaction channels, charged particles and
neutrons were detected by the EUCLIDES silicon apparatus \cite{EUCLIDES} and the Neutron Wall array \cite{nw1,nw2}, respectively.Data were recorded by the GALILEO data acquisition system which was designed for the GALILEO-EUCLIDES-NWALL Experiment \cite{Berti2015}. More details of the experimental setup and data
analysis can be found in Ref. \cite{130ba-isomer} and a forthcoming paper \cite{130ba-gs}.

\begin{figure*}[ht]
\hskip  1. cm
\rotatebox{-0} {\scalebox{.85}{\includegraphics{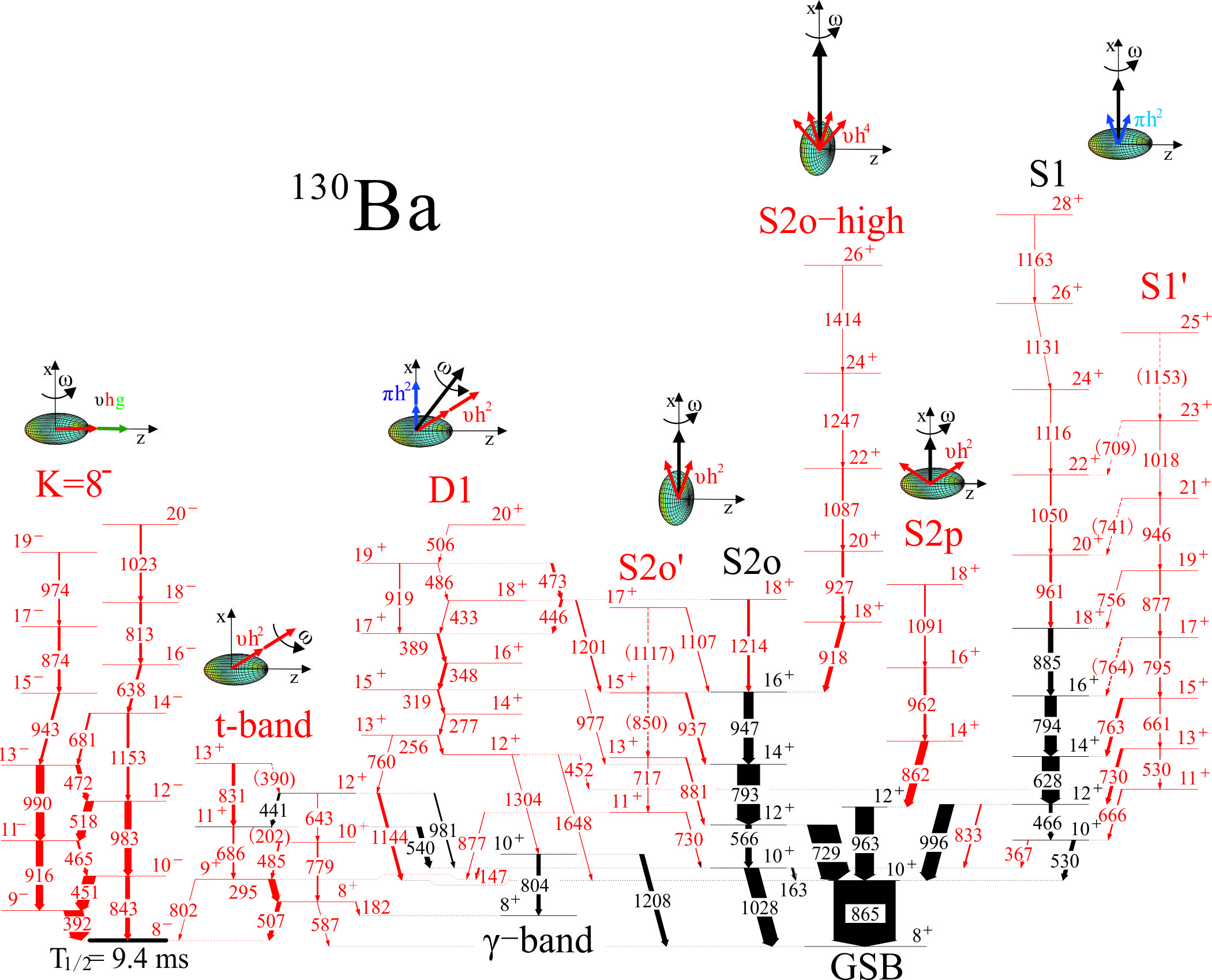}}}
\vskip -.0 cm
\caption {Partial level scheme of $^{130}$Ba showing the newly observed bands feeding through the $8^+$ GSB member at 2396 keV. The assigned configuration and corresponding shape are schematically sketched above each band. The rotation axes and the orientation of the angular momenta of the nucleons involved in each configuration are also indicated. Known (new) levels and transitions are drawn with black (red) colors, respectively. }
\label{fig1} 
\end{figure*}

\begin{figure}[]
\includegraphics[angle=0,width=8.5cm]{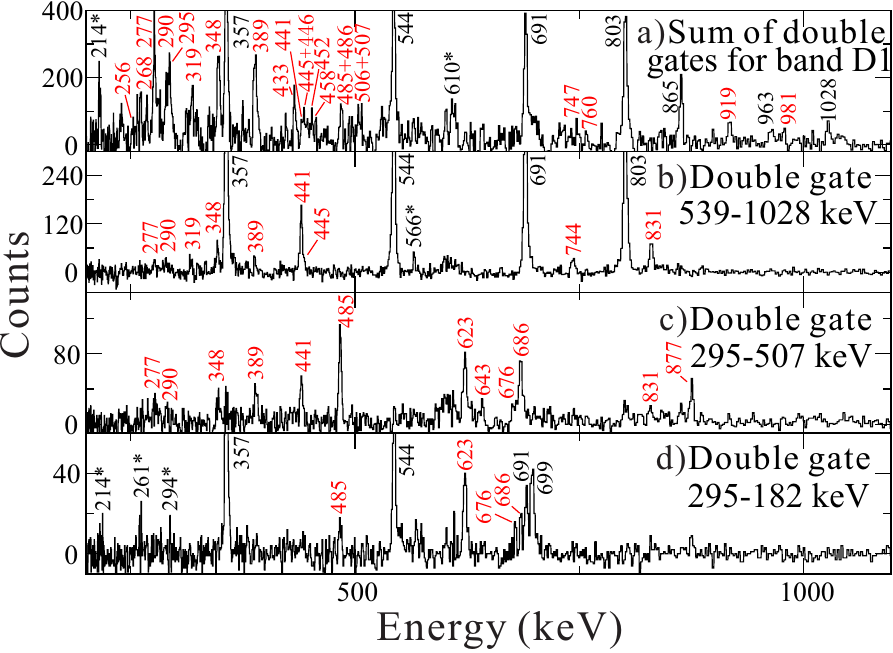}
\caption{(Color online) Typical double-gated spectra for bands $D1$ and $t$-band. The gates were placed on selected transitions labelled
in each panel, while panel a) is the sum of double-gated spectra using all the transitions in band $D1$ except the 277-keV transition. Transitions belong to the new structure are marked in red, while those marked with asterisks are identified contaminants.}
\label{fig2}
\end{figure}

In the present work, five new bands with positive parity were identified and the previously known bands built on the $10^+$ states were extended at higher spins, as shown in Fig. \ref{fig1}. 
Representative double-gated spectra are shown in Fig. \ref{fig2} and \cite{SM}. The $11^+$ and $12^+$ levels of the $t$-band were previously
established by means of the decaying 441-, 540- and 981-keV transitions  \cite{Ka14}, whereas the levels with spins $8^+$, $9^+$,
$10^+$ and $13^+$ are new. Several new transitions linking the $t$-band to low-lying states are identified.
Band $D1$ built on the $12^+$ state is completely new. It decays directly and through
intermediate states, which will be published in a forthcoming paper \cite{130ba-gs}, to the $t$-band, $\gamma$-band, ground state band ($GSB$), and to bands $S1$ and $S2o$. Interestingly, there is an accidental degeneracy between the $18^+$ states of band $D1$ and $S2o$ lying 14 keV apart, which gives rise to the 446-, 473- and 1201-keV connecting transitions. Also the $15^+$ states of the bands $D1$ and $S2o$ are separated by only 40 keV, which can explain the observation of the 977-keV connecting transition. These accidental degeneracies are important in understanding the connecting transitions between the bands $D1$ and $S2o$, which are interpreted as based on prolate and oblate shapes, respectively (see the following discussion subsection). For the bands populated with sufficient intensity, we extracted the mixing ratios $\delta$ from the analysis of the angular distributions, and deduced the $B(M1)/B(E2)$ and $B(E2)_{out}/B(E2)_{in}$ ratios of reduced transition probabilities (see the supplemental material \cite{SM}). 

The bands $S1$ and $S2o$ previously reported in Ref. \cite{Ka14} are confirmed only up to spin $18^+$ and $16^+$, respectively. The previously reported 927-keV transition of band $S1$, as well as 1027- and 1040-keV transitions of band $S2o$ are not observed and therefore not confirmed by the present data. Five transitions of 961, 1050, 1117, 1130 and 1163 keV in band $S1$, and four transitions of 927, 1087, 1247 and 1414 keV composing the band $S2o$-high, which decays to band $S2o$ through the 918-keV transition, are newly identified at high spins. 
The odd-spin bands $S1'$ and $S2o'$, as well as their connecting transitions to the even-spin partners are completely new. Detailed experimental information on the transitions of bands $S1$, $S1'$, $S2o$ and $S2o'$ will be published in a forthcoming paper \cite{130ba-gs}. 

We also identified the high-spin part of the ground-state band, $S2p$, consisting of the 962-, 1091-keV transitions, which decays to the $12^+$ state of $GSB$ via the 862-keV transition. Three new levels with spins $15^-$, $17^-$ and $19^-$ on top of the
odd-spin cascade built on the $8^-$ isomer are newly identified, which can be the signature partner of the even-spin cascade reported previously  \cite{130ba-isomer}.

\section{Discussion}

In order to understand the nature of the observed positive-parity bands we performed a detailed theoretical investigation, employing several theoretical models. It appears that the medium-spin states of $^{130}$Ba constitute one of the most complete sets of bands built on different shapes rotating around axes with different orientations, offering thus a fertile field of investigation of multi-quasiparticle collective excitations. In fact, we already had indications that $^{130}$Ba, a nucleus with a pronounced $\gamma$-softness at low spin, can acquire a stable axial shape in two-quasiparticle configurations, as in the $\nu h_{11/2}^{-1} g_{7/2}^{-1}$ configuration assigned to the $8^-$ isomer \cite{130ba-isomer}.  

 In order to understand the observed structures and to assign them configurations, we first extracted the effective moments of inertia (MOI) from the slope of the linear part of the $I-\omega$ curves, which were subsequently used to determine the single-particle aligned angular momenta $i_x$ following the procedure of Ref. \cite{bf1979,bengtsson1986} (see the supplemental material \cite{SM}).

The first important result of the present work is the observation of the odd-spin bands $S1'$ and $S2o'$, of the even-spin bands $S1$ and $S2o$. There is very scarce experimental information on such odd-spin states decaying to the states of the $S$-bands, typically only a few levels being observed \cite{128xe-orce,126ba-ward,128ba,132ba,134ce-petrache,136nd-lv,138nd-petrache}. 
Their interpretation is often handwaving, invoking the $\gamma$-band built on the $S$-bands, or is completely missing. These weakly populated odd-spin bands are however essential for the characterization of the configurations on which the $S$-bands are built, in particular the $K$-composition and the nuclear shape. To our knowledge, the present bands $S1'$ and $S2o'$ are among the most extended cascades composed of odd-spin states decaying to the $S$-bands observed in the $A=130$ mass region. We therefore paid special attention to the analysis of these bands and tried to globally understand the ensemble of bands denoted 
$S1$, $S1'$, $S2o$, $S2o'$ and $S2p$ in Fig. \ref{fig1}.

The second important result of the present work is the observation of the $t$-band, which is for the first time identified in the $A=130$ mass region. 
The assigned $\nu h_{11/2}^{2}$ configuration of the $t$-band is composed of the two FAL quasineutron orbitals, which leads to rotation around a tilted axis, as schematically drawn in Fig. \ref{fig1}. No such band has been observed in the neighboring even-even nuclei, in particular in the isotone $^{132}$Ce, which has a similar level scheme \cite{132ce-paul}. This structure can, however, be compared with the $K^{\pi}=23/2^+$ band in $^{129}$Ba \cite{129ba-byrne}, as well as with the recently reported high-$K$ band in $^{127}$Xe \cite{127xe-Chakraborty}, both being built on isomeric states that support their high-$K$ assignments. In $^{130}$Ba the band-head of the $t$-band is not isomeric, but has a significant decay branch to the $K^{\pi} = 8^-$ isomer, indicating its high-$K$ character. Such $t$-bands, built on $\nu i^2_{13/2}$ configurations, have also been identified in the A $\approx$180 region \cite{Walker2016Phys.Scr013010}. A specific comparison can be made with $^{180}$W \cite{180W-walker}, in which a band built on a $8^+$ state interpreted as having a $\nu i_{13/2}^2$ configuration has a prompt decay branch to a $K^{\pi}= 8^-$ isomer, like in the case of $^{130}$Ba. We therefore suggest that the newly identified $t$-band of $^{130}$Ba provides the first evidence of a two-quasiparticle $t$-band in the $A = 130$ mass region.

We also identified band $D1$, composed of two degenerate signature partner cascades connected by intense $\Delta I = 1$ transitions,
which strongly suggests the presence in its configuration of two high-$\Omega$ neutron orbitals, which can be the same
as in the $t$-band ($\nu h_{11/2}^{2}$ FAL) and two low-$\Omega$  $\pi h_{11/2}$ protons. We therefore assign the $\pi h^2_{11/2} \otimes \nu h_{11/2}^{2}$FAL configuration to band $D1$, which is obtained by coupling the prolate configurations $\pi h^2_{11/2}$ 
assigned to band $S1$ and $\nu h_{11/2}^{2}$  assigned to the $t$-band. 

\subsection{TRS calculations}

The bands $S1$ and $S2o$ have been reported previously in Ref. \cite{130ba-sun}, and interpreted, based on cranking calculations, as two-quasiparticle proton and neutron aligned
bands, respectively. Similar conclusions have been also drawn in other theoretical papers devoted to the structure of the Xe, Ba, Ce nuclei \cite{rohozinski1977,faessler1985,hammaren1986,wyss1988,wyss1989,granderath1996}.

In order to check this interpretation, we first performed  TRS calculations \cite{trs1988}, to understand which are the favored two-quasiparticle alignments (see supplemental material \cite{SM}). The TRS calculations confirm that a pair of $h_{11/2}$ neutrons decouples first and aligns to the rotation axis at a rotational frequency of 0.35 MeV/$\hbar$, driving the nuclear shape towards a near-oblate shape with $(\beta_2, \gamma)\approx(0.17, -45^{\circ})$. Coexisting are near-prolate minima with $(\beta_2, \gamma)\approx(0.20, +15^{\circ})$ and  $(\beta_2, \gamma)\approx(0.20, -15^{\circ})$ which support the bands S1, t, 8$^-$ and D1. The assignment of S2p will be discussed below.

\subsection {TAC calculations}

\begin{figure}[h]
\hskip -0.cm
\includegraphics[angle=0,width=8.5cm]{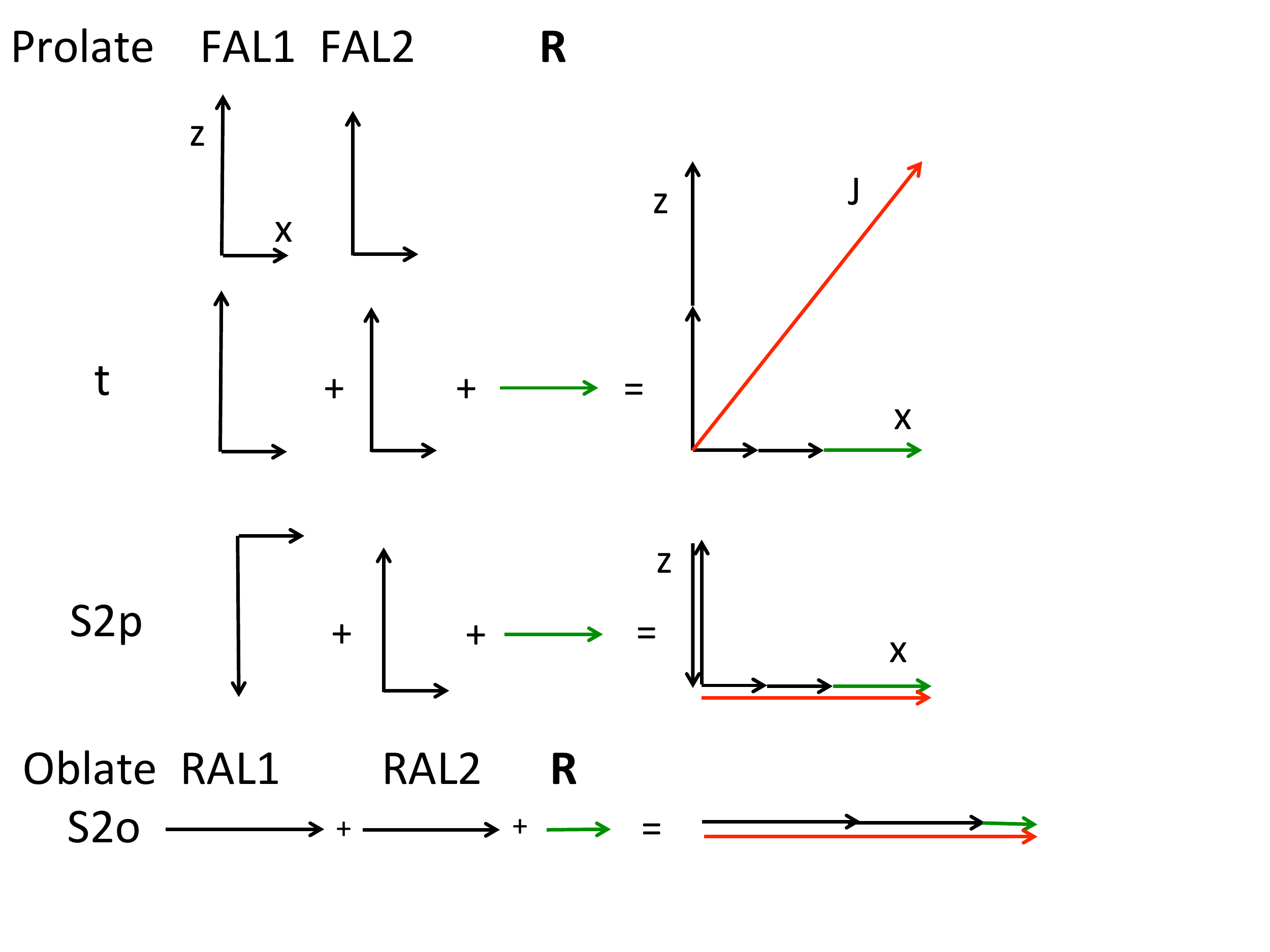}
\vskip-0.cm
\caption{(Color online) Schematic representation illustrating the configurations of the $t$-band and the two $S$-bands $S2p$  and $S2o$ which are based on the same configuration but different shapes. For prolate shape, the angular momentum components of the two Fermi-aligned (FAL1 and FAL2) orbits active in the $t$-band are larger on the symmetry axis ($z$) than on the rotation axis ($x$). The angular momentum of the core $R$ is perpendicular to the symmetry axis. For oblate shape, the angular momentum components of the two rotation-aligned (RAL1 and RAL2) orbits active in band  $S2o$ are along the rotation axis. Color code
of the angular momentum vectors:  Black - quasineutrons $\vec j$,  Green - collective  $\vec R$, Red - total $\vec J$.}
\label{fig3}
\end{figure}

\begin{figure}[h]
\hskip -0.cm
\includegraphics[angle=0,width=6.5cm]{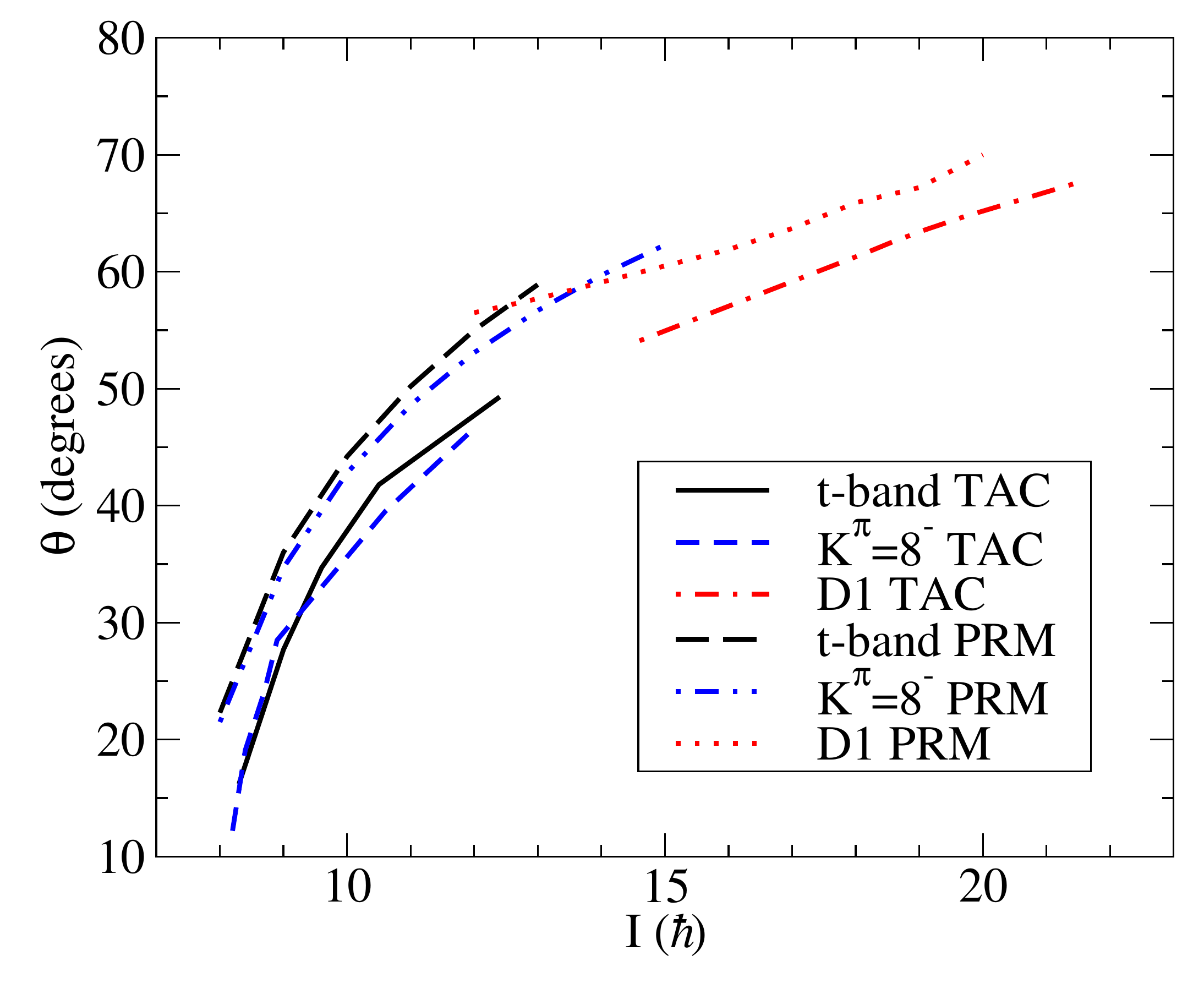}
\vskip-0.cm
\caption{(Color online) Angle $\theta$ of the angular momentum vector $\vec J$ with respect to the symmetry axis 3   calculated using the TAC and PRM models for the configurations assigned to the $t$-band, $K^{\pi}=8^-$ and $D1$ bands. }
\label{fig4}
\end{figure}

In order to investigate the orientation of the rotation axis, the components of the angular momenta on the three axes of the intrinsic system and the transition probabilities we carried out TAC calculations \cite{fra-2000} for the two axial deformations  $\varepsilon_2=0.20$ and -0.19, and $\varepsilon_4=0.02$, which are close to the coexisting minima found in the TRS calculations and the deformations used in the PRM and PSM calculations. 
 A schematic drawing of the different angular momentum coupling of the $\nu h_{11/2}^{2}$ quasineutron pair in the FAL $t$-band, and the RAL bands $S2o$ and $S2p$ with oblate and prolate shapes, respectively, is shown in Fig. \ref{fig3}. 
The angles of the total angular momentum $I$ with respect to the symmetry  axis (z) for the $t$-band and the $K^{\pi}=8^-$ band have similar behavior. They are much smaller than that of band $D1$ (see Fig. \ref{fig4}),
because the  low-$\Omega$ quasiproton pair drives $\vec J$ toward 90$^\circ$. The energies and  $B(M1)/B(E2)$ ratios of reduced transition probabilities are shown in Figs. \ref{fig6} and  \ref{fig7} of the following subsection on PSM calculations. One can see a good agreement with the measured values of all three bands, $K^{\pi}=8^-$, $t$-band, and band $D1$.

The two FAL can be alternatively combined to form the configuration denoted $S2p$ in Fig. \ref{fig3}, which is observed in $^{180}$W and $^{182}$Os 
 \cite{180W-walker,182os-1995} as an up-bend of the even-$I$ yrast sequence. In analogy, we suggest assigning a configuration
 of the  $S2p$ type to the band that crosses the $GSB$ at $I$=12 in Fig. \ref{fig1}, which is based on the following observations.
 The gain of angular momentum  at the crossing with the $GSB$ indicates that  it must be caused by the alignment 
 of a pair of $h_{11/2}$ quasiparticles. At oblate shape the RAL quasineutrons $e, f, g$ have increasing energy and signatures 
 $\alpha = -1/2, 1/2, -1/2$, respectively. The band $S2o$ is  assigned to $\nu [ef$]  and  $S2o$'  to $\nu [eg$]. We have discarded
 the possible assignment of $S2p$ to $\nu [fg$] because there are no connecting transitions to $S2o'$ like the ones observed
  from $S2o'$ to $S2o$.
 At prolate shape the analogous structure appears for the RAL quasiprotons. We have discarded the  possible
 assignment $\pi [fg$] because there are no connecting transitions observed from $S2p$ to $S1'$. Ruling out the two
 possibilities leaves the combination of the two FAL  quasineutrons to $S2p$ as alternative. 
 
For the $\nu h_{11/2}^2$ configuration the TRS calculations show a minimum at $\gamma \approx -60^\circ$ and
 a kind of plateau around $\gamma=0^\circ$. $^{130}$Ba is very $\gamma$-soft (close to the O(6) limit \cite{kirch}), and large-scale collective motion
 in the $\gamma$-degree of freedom is expected. The collective wave functions for a Bohr Hamiltonian with a qualitatively
 similar potential are shown in Ref. \cite{caprio} (Fig. 8, case $\chi=50$, change $\gamma\rightarrow60^\circ-\gamma$. The presence of such a state at 1179 keV was demonstrated in Ref. \cite{kirch}.)
 The collective ground state  ($S2o$) is centered near oblate shape.
 The first exited spin-zero state ($\gamma\gamma_0$) has a node and two maxima, one on the oblate side 
 and one (with larger probability) 
 near $\gamma=30^\circ$.  The calculation assumes  a constant mass parameter. It is possible that a $\gamma$-dependent mass parameter gives more weight to the prolate side. With increasing $\gamma$ the FAL states in  $S2p$ develop toward RAL, i. e.
 the $z$-component decreases and the $x$-component increases, which means that the total $x$-component increases
 while the $z$-component remains zero. We keep the label $S2p$ because the structure is the same as for prolate shape.   

\subsection{PRM calculations}

The active nucleon configurations that can be assigned to the bands observed in $^{130}$Ba have been calculated using the configuration-fixed constrained triaxial covariant density functional theory (CDFT) framework \cite{meng2016} and those calculated using PC-PK1 effective interaction \cite{Zhao2010Phys.Rev.C054319} are given in Table \ref{tab3}. With the obtained deformation parameters, quantal PRM calculations \cite{135nd-prm,streck2018, Ch18} have been carried out to describe the experimental energy spectra and electromagnetic transition probabilities. Both the  CDFT and PRM calculations were performed without pairing.  

\begin{figure}[]
\includegraphics[angle=0,width=6.5cm]{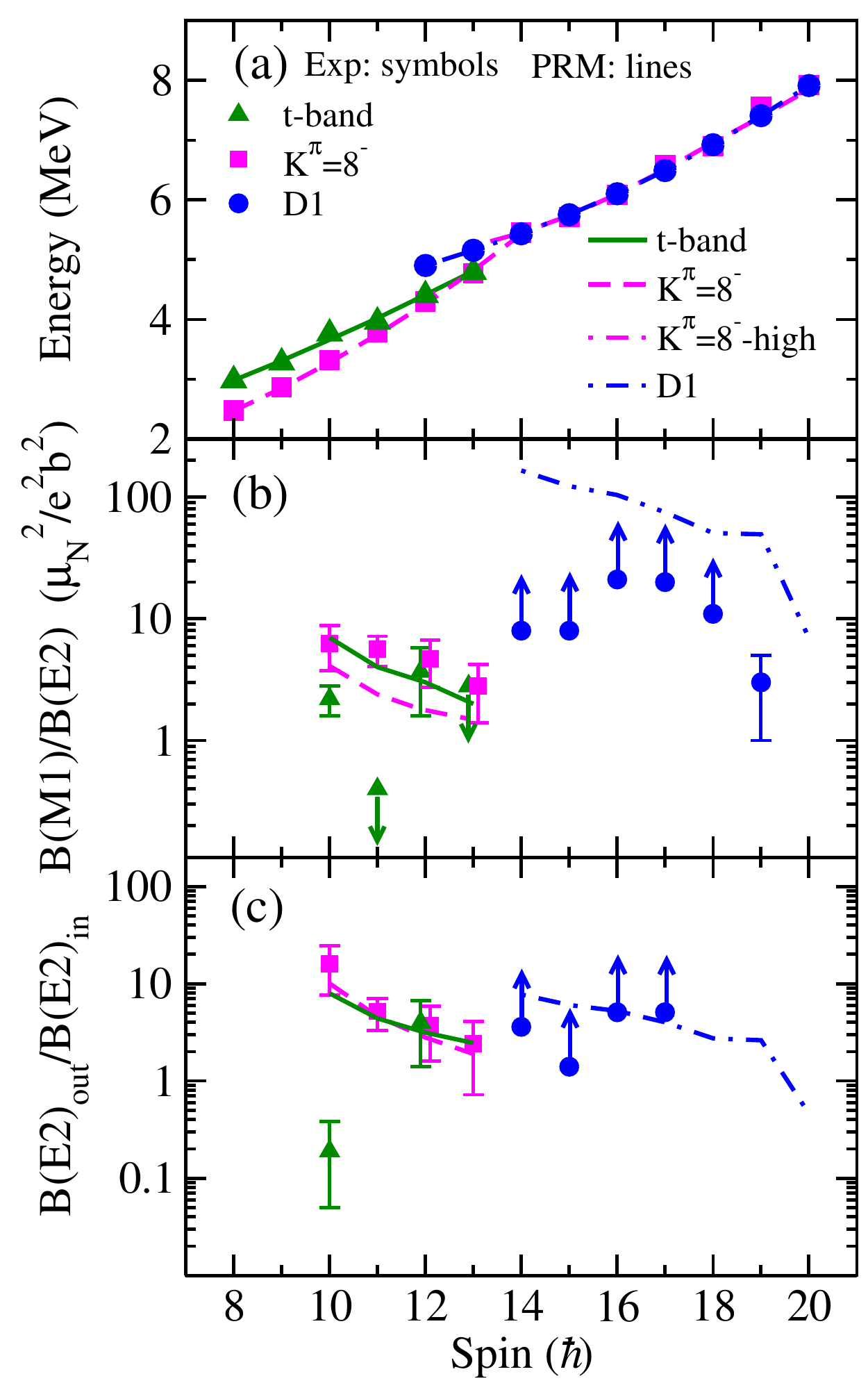}
\caption{(Color online) Comparison between experimental and PRM results for the bands $D1$, $K^{\pi}=8^-$ and $t$-band assuming prolate shapes.}
\label{fig5}
\end{figure}

The used moments of inertia $\mathcal{J}_0$ and Coriolis attenuation factors $\xi$ in the PRM calculations are also listed in Table \ref{tab3}. The obtained calculated results in comparison with the experimental data are shown in Fig. \ref{fig5}. It is seen that the experimental energy spectra, $B(M1)/B(E2)$, as well as $B(E2)_{\rm{out}}/B(E2)_{\rm{in}}$ of bands $K^\pi=8^-$, $t$-band, and $D1$ are described reasonably well by the PRM. This gives strong support for the configurations assigned to these three bands. 

One notes that the experimental $B(M1)/B(E2)$ of the $K^\pi=8^-$ band and the $t$-band are similar. The $B(M1)/B(E2)$ values of the $t$-band are overall a bit larger that those of the $K^\pi=8^-$ band, because the calculated $B(M1)$ values of the $t$-band are larger that those of the $K^\pi=8^-$ band. This can be attributed to the different effective $g$-factors $(g_\nu-g_R)$ (with $g_R=Z/A$) of the $\nu h_{11/2}$ and $\nu g_{7/2}$ orbitals. For the $\nu h_{11/2}$ orbital, the $(g_\nu-g_R)=-0.64$, while for the $\nu g_{7/2}$ orbital $(g_\nu-g_R)=-0.18$. The two quasineutrons in the two bands have a large angular momentum component $j_3$ to generate the band-head spin $I = 8$ (see Figs. \ref{fig1} and \ref{fig5}). Hence, the $M1$ operator and consequently the $B(M1)$ values of the $t$-band, which are proportional to the square of the magnetic moment, are larger than those of the $K^\pi=8^-$ band. 

Moreover, the $B(E2)$ values are determined by the Clebsch-Gordan coefficients $\langle I_i K 20 | I_f K \rangle$ for a band with good $K$ \cite{BM}. When the spin $I_i$ is small, the Clebsch-Gordan coefficients for $I_f=I_i-1$ is larger than that for $I_f=I_i-2$. As a consequence, the $B(E2)_{\rm{out}}$ is larger than the $B(E2)_{\rm{in}}$. This feature is seen in Fig. \ref{fig5} for the $t$-band and the $K^\pi=8^-$ band. Furthermore, due to the same nominal $K=8$ value in the two bands, their $B(E2)_{\rm{out}}/B(E2)_{\rm{in}}$ values are also similar, though there is more $K$-mixing in the $t$-band. 

Fig. \ref{fig4} shows the angle of the total angular momentum with the (near-) symmetry axis 3. It is obtained as 
$\cos \theta=\sqrt{\langle J_3^2\rangle}/(I+1/2)$.
The results are in good agreement with the TAC calculations. 

\begin{table}
  \caption{The deformation
           parameters $\beta$ and $\gamma$, moments of inertia  $\mathcal{J}_0$ (in units of $\hbar^2/\textrm{MeV}$), $\xi$ parameters, and their corresponding
           active nucleon configurations in the configuration-fixed
           constrained triaxial covariant density functional theory calculations.}\label{tab3}
    \begin{tabular}{cccc}
      \hline
      \hline
 $(\beta, \gamma)$ & $\mathcal{J}_0$ & $\xi$ & $\rm Active~nucleon~configuration$ \\
                              & ($\hbar^2/$MeV)  &          &                                                        \\
      \hline
$(0.23, 13.9^\circ)$ & &&Nucleons paired \\
$(0.23, 11.7^\circ)$ & 21 & 1.00 & $\nu[514]9/2^-[404]7/2^+$ \\
$(0.21, 16.8^\circ)$ & 22 & 1.00 & $\nu h_{11/2}^{-2}$ \\
$(0.23, 24.2^\circ)$ & 18 & 0.96 & $\nu h_{11/2}^{-2} \otimes \pi h_{11/2}^2$ \\
$(0.25, 30.1^\circ)$ & 24 & 0.95 & $\nu[514]9/2^-[404]7/2^+ \otimes \pi h_{11/2}^2$\\

\hline
\end{tabular}
\end{table}

\subsection{PSM calculations}

PSM \cite{hara-sun,Sun-2016} has been successfully applied for studying the structure of high-spin states,
such as tilted rotation in the A$\approx180$ mass region \cite{Sheikh1998Phys.Rev.R26}, multi-quasiparticle configurations \cite{LJ-2014,YX-2011} and multiple dipole bands \cite{costel-1996,costel-1997}.
Angular-momentum projection is performed for each $\textit{K}$ configuration and the mixing
among states with different $\textit{K}$ values is calculated by diagonalizing
the shell model Hamiltonian in the projected basis.

In the PSM, one first determines a deformed basis for a calculation.
We adopted the quadrupole $\varepsilon_{2}=0.22$ and hexadecapole
deformations $\varepsilon_{4}=0.02$ suggested in Ref. \cite{moller2016} for
$^{130}$Ba and assumed axial symmetry. The monopole-pairing strength is taken to
be $G_{M}=[20.82\mp13.58(N-Z)/A]/A$, for neutrons and protons, respectively.
The quadrupole-pairing strength $G_{Q}$ is assumed
to be proportional to $G_{M}$, with the proportionality constant $0.18$ for $^{130}$Ba.
For the valence single-particle space,
we include three major shells, N $= 3, 4, 5$, for both neutrons and protons.
For oblate deformation, all model parameters are the same except for the
quadrupole deformation which is $\varepsilon_{2}=-0.18$.

The band diagrams which display the angular-momentum-projected energies versus spins of rotational bands before configuration
mixing \cite{hara-sun} are shown in the supplemental material \cite{SM} for prolate and oblate shapes. 

\begin{figure}[]
\hskip -.cm
\includegraphics[width=3.6in]{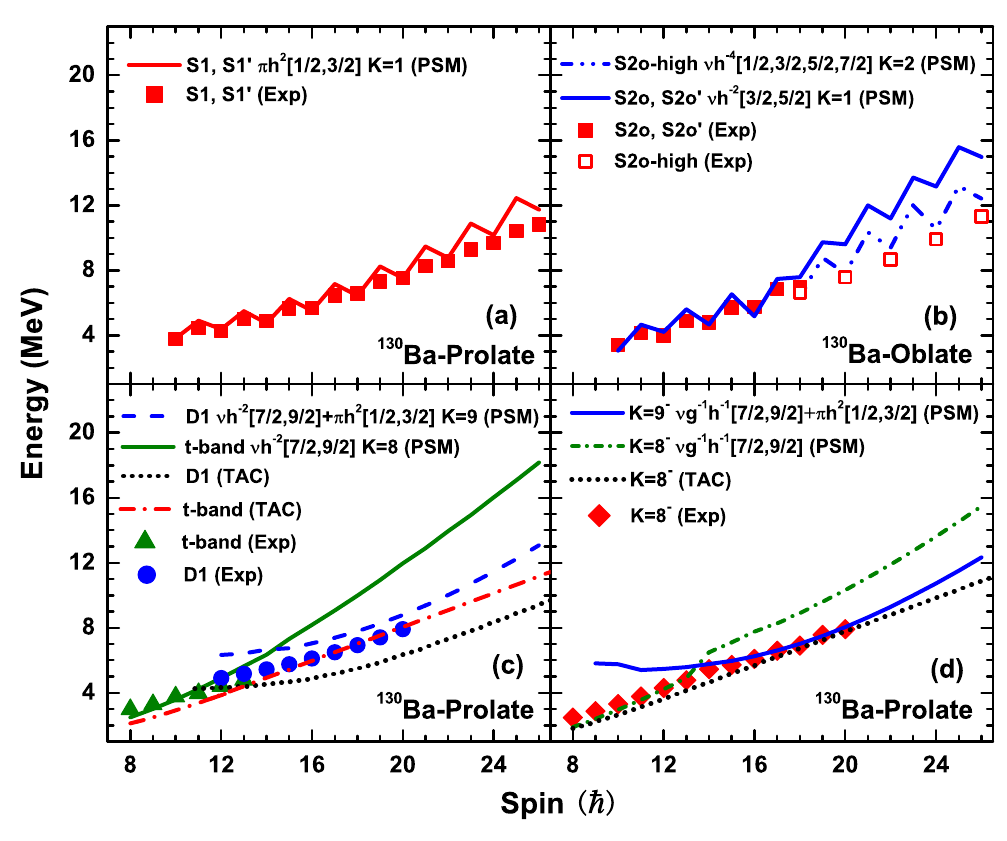}
\caption{(Color online) Comparison of the  calculated PSM energies with available data for $^{130}$Ba. For the $t$-band, $D1$ and $K^{\pi}=8^-$ bands, also the calculated TAC energies are included in panes (c) and (d).}
\label{fig6} 
\end{figure}

The PSM results after configuration mixing are compared with the experimental data in Fig. \ref{fig6}. The theoretical
results are in good agreement with the available experimental data for all rotational bands considered in the present calculation. 
 From Fig. \ref{fig6}(a), we can see that band $S1$ is well reproduced by the 2-quasiparticle proton configuration that is 
 dominated by the $\pi h^{2}[1/2,3/2]$, $K^{\pi}=1^{+}$, component. However, the calculated staggering is larger than 
 the experimental one above spin $I=20$.
 The comparison of the calculated $(S2o,S2o')$ and $S2o$-high bands with the available data is shown in Fig. \ref{fig6}(b).
A very good agreement with the experimental data is obtained over the entire observed spin range, which strongly supports the oblate 2-quasiparticle neutron configuration assigned to bands $(S2o, S2o')$, which is crossed by the 4-qp neutron configuration assigned to band $S2o$-high.

The energies of the prolate configurations assigned to band $D1$, $t$-band, and $K^{\pi}=8^-$ band shown in Fig. \ref{fig6}(c) increase smoothly with spin, without odd-even staggering. The 2-quasiparticle neutron configuration dominated by the $\nu h^{2}[7/2,9/2], K=8$ component is assigned to $t$-band. It reproduces well the experimental data. The 4-quasiparticle configuration dominated by the $\nu h^{-2}[7/2,9/2]+\pi h^{2}[1/2,3/2], K=9$  component is assigned to band $D1$. It is in qualitative agreement with the experimental data, with a larger deviation towards the band-head. The difference between calculated and experimental energies can be reduced by taking into account the triaxial deformation, but this is beyond the scope of the present work and will be addressed in a future publication with calculations using the triaxial projected shell model (TPSM) \cite{sheikh-TPSM}.  

The calculated negative-parity band built on the $8^-$ isomer is compared with the experimental data in Fig. \ref{fig6}(d).
The 2-quasiparticle configuration dominated by the $\nu g^{1}h^{1}[7/2,9/2], K^{\pi}=8^-$ and the 4-quasiparticle configuration dominated by the $\nu g^{1}h^{1}[7/2,9/2]\otimes\pi h^{2}[1/2,3/2], K^{\pi}=9^{-}$ component reproduce very well the experimental data in the spin ranges $I=8-14$ and $I=15-20$, respectively. 

\begin{figure}[]
\centering
\includegraphics[width=3.in]{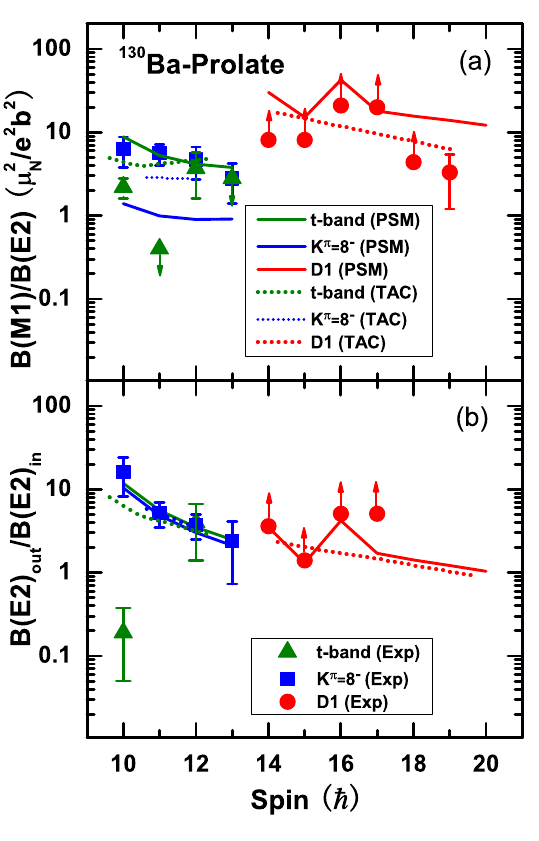}
\caption{(Color online) Comparison of the experimental and calculated $B(M1)/B(E2)$ and  $B(E2)_{\textrm{out}}/B(E2)_{\textrm{in}}$ ratios for $t$-band, $D1$ and $K^{\pi}=8^{-}$ bands using the PSM and TAC models, and employing the same deformation parameters ($\varepsilon_2=0.20, \gamma=0^{\circ})$.}
\label{fig7} 
\end{figure}

In the PSM calculations, the $B(E2)$, $B(M1)$ and $g$-factors are evaluated using the (final)
shell model wave functions, as illustrated in the early PSM work of Ref. \cite{hara-sun,Sun-psm}. We
emphasize that the $g$-factor in the PSM is computed directly from the many-body wave function without
a semiclassical separation of the collective and the single-particle parts.

The calculated $B(M1)/B(E2)$ and $B(E2)_{\textrm{out}}/B(E2)_{\textrm{in}}$ for $t$-band,
$K^{\pi}=8^{-}$ and $D1$ bands are compared with available
experimental data in Fig. \ref{fig7}. For $t$-band and band $D1$, the experimental $B(M1)/B(E2)$ data are well reproduced. However, the calculated results of the $K^{\pi}=8^{-}$ band are lower than the experimental data,
due to the smaller calculated $B(M1)$ values, while those for the $K^{\pi}=8^+$ band are in agreement with the experimental values. However, we need to be cautious, because the $8^+$ band ($t$-band) is significantly non-yrast, and mixings due to accidental degeneracies could be involved. For the $B(E2)_{\textrm{out}}/B(E2)_{\textrm{in}}$, the PSM results are in good agreement with the data.

Summarizing, the present work reports the first observation of a two-quasiparticle $t$-band in the $A=130$ mass region, as well as the odd-spin partners of the $S$-bands built on prolate 2qp-proton and oblate 2qp-neutron configurations. Extended calculations using several theoretical models converge in a coherent interpretation of the observed bands, which represent one of the best examples of shape coexistence and exotic rotations that a $\gamma$-soft nucleus can exhibit at medium and high spins.

This research is supported by the National Natural Science Foundation of China (Grants No. U1832139, Grants No. 11505242, No. 11305220, No. U1732139, No. 11775274 and No. 11575255), by the Minister of Europe and Foreign Affairs, Partnership Hubert Curien, project Cai YuanPei 2018, n. 41458XH, by the Natural Sciences and Engineering Research Council of Canada, by the National Research, Development and Innovation Fund of Hungary (Project no. K128947), by STFC funding (grant ST/P005314/1), by the GINOP-2.3.3-15-2016-00034, National Research, Development and Innovation Office NKFIH, Contract No. PD 124717, by the Polish National Science Centre (NCN) Grant No. 2013/10/M/ST2/00427, and by the National Research Foundation of South Africa, GUN: 93531, 109134. The work of Q. B. C. is supported by Deutsche Forschungsgemeinschaft (DFG) and National Natural Science Foundation
of China (NSFC) through funds provided to the Sino-German CRC 110 ``Symmetries and the Emergence of Structure in QCD".

\bibliography{Article-t-band}

\begin{thebibliography}{56}%
\makeatletter
\providecommand \@ifxundefined [1]{%
 \@ifx{#1\undefined}
}%
\providecommand \@ifnum [1]{%
 \ifnum #1\expandafter \@firstoftwo
 \else \expandafter \@secondoftwo
 \fi
}%
\providecommand \@ifx [1]{%
 \ifx #1\expandafter \@firstoftwo
 \else \expandafter \@secondoftwo
 \fi
}%
\providecommand \natexlab [1]{#1}%
\providecommand \enquote  [1]{``#1''}%
\providecommand \bibnamefont  [1]{#1}%
\providecommand \bibfnamefont [1]{#1}%
\providecommand \citenamefont [1]{#1}%
\providecommand \href@noop [0]{\@secondoftwo}%
\providecommand \href [0]{\begingroup \@sanitize@url \@href}%
\providecommand \@href[1]{\@@startlink{#1}\@@href}%
\providecommand \@@href[1]{\endgroup#1\@@endlink}%
\providecommand \@sanitize@url [0]{\catcode `\\12\catcode `\$12\catcode
  `\&12\catcode `\#12\catcode `\^12\catcode `\_12\catcode `\%12\relax}%
\providecommand \@@startlink[1]{}%
\providecommand \@@endlink[0]{}%
\providecommand \url  [0]{\begingroup\@sanitize@url \@url }%
\providecommand \@url [1]{\endgroup\@href {#1}{\urlprefix }}%
\providecommand \urlprefix  [0]{URL }%
\providecommand \Eprint [0]{\href }%
\providecommand \doibase [0]{http://dx.doi.org/}%
\providecommand \selectlanguage [0]{\@gobble}%
\providecommand \bibinfo  [0]{\@secondoftwo}%
\providecommand \bibfield  [0]{\@secondoftwo}%
\providecommand \translation [1]{[#1]}%
\providecommand \BibitemOpen [0]{}%
\providecommand \bibitemStop [0]{}%
\providecommand \bibitemNoStop [0]{.\EOS\space}%
\providecommand \EOS [0]{\spacefactor3000\relax}%
\providecommand \BibitemShut  [1]{\csname bibitem#1\endcsname}%
\let\auto@bib@innerbib\@empty
\bibitem [{\citenamefont {Qiang~{\it et al.}}(2019)}]{130ba-isomer}%
  \BibitemOpen
  \bibfield  {author} {\bibinfo {author} {\bibfnamefont {Y.~H.}\ \bibnamefont
  {Qiang~{\it et al.}}},\ }\href {\doibase 10.1103/PhysRevC.99.014307}
  {\bibfield  {journal} {\bibinfo  {journal} {Phys.Rev. C}\ }\textbf {\bibinfo
  {volume} {99}},\ \bibinfo {pages} {014307} (\bibinfo {year}
  {2019})}\BibitemShut {NoStop}%
\bibitem [{\citenamefont {Wyss}\ \emph
  {et~al.}(1988{\natexlab{a}})\citenamefont {Wyss}, \citenamefont {Nyberg},
  \citenamefont {Johnson}, \citenamefont {Bengtsson},\ and\ \citenamefont
  {Nazarewicz}}]{trs1988}%
  \BibitemOpen
  \bibfield  {author} {\bibinfo {author} {\bibfnamefont {R.~A.}\ \bibnamefont
  {Wyss}}, \bibinfo {author} {\bibfnamefont {J.}~\bibnamefont {Nyberg}},
  \bibinfo {author} {\bibfnamefont {A.}~\bibnamefont {Johnson}}, \bibinfo
  {author} {\bibfnamefont {R.}~\bibnamefont {Bengtsson}}, \ and\ \bibinfo
  {author} {\bibfnamefont {W.}~\bibnamefont {Nazarewicz}},\ }\href {\doibase
  10.1016/0370-2693(88)91422-0} {\bibfield  {journal} {\bibinfo  {journal}
  {Phys. Lett. B}\ }\textbf {\bibinfo {volume} {215}},\ \bibinfo {pages} {211}
  (\bibinfo {year} {1988}{\natexlab{a}})}\BibitemShut {NoStop}%
\bibitem [{\citenamefont {Petrache}\ \emph {et~al.}(2012)\citenamefont
  {Petrache}, \citenamefont {Frauendorf}, \citenamefont {Matsuzaki},
  \citenamefont {Leguillon}, \citenamefont {Zerrouki}, \citenamefont {Lunardi},
  \citenamefont {Bazzacco}, \citenamefont {Ur}, \citenamefont {Farnea},
  \citenamefont {Rossi~Alvarez}, \citenamefont {Venturelli},\ and\
  \citenamefont {de~Angelis}}]{138nd-low}%
  \BibitemOpen
  \bibfield  {author} {\bibinfo {author} {\bibfnamefont {C.~M.}\ \bibnamefont
  {Petrache}}, \bibinfo {author} {\bibfnamefont {S.}~\bibnamefont
  {Frauendorf}}, \bibinfo {author} {\bibfnamefont {M.}~\bibnamefont
  {Matsuzaki}}, \bibinfo {author} {\bibfnamefont {R.}~\bibnamefont
  {Leguillon}}, \bibinfo {author} {\bibfnamefont {T.}~\bibnamefont {Zerrouki}},
  \bibinfo {author} {\bibfnamefont {S.}~\bibnamefont {Lunardi}}, \bibinfo
  {author} {\bibfnamefont {D.}~\bibnamefont {Bazzacco}}, \bibinfo {author}
  {\bibfnamefont {C.~A.}\ \bibnamefont {Ur}}, \bibinfo {author} {\bibfnamefont
  {E.}~\bibnamefont {Farnea}}, \bibinfo {author} {\bibfnamefont
  {C.}~\bibnamefont {Rossi~Alvarez}}, \bibinfo {author} {\bibfnamefont
  {R.}~\bibnamefont {Venturelli}}, \ and\ \bibinfo {author} {\bibfnamefont
  {G.}~\bibnamefont {de~Angelis}},\ }\href {\doibase
  10.1103/PhysRevC.86.044321} {\bibfield  {journal} {\bibinfo  {journal} {Phys.
  Rev. C}\ }\textbf {\bibinfo {volume} {86}},\ \bibinfo {pages} {044321}
  (\bibinfo {year} {2012})}\BibitemShut {NoStop}%
\bibitem [{\citenamefont {Dracoulis}\ \emph {et~al.}(2016)\citenamefont
  {Dracoulis}, \citenamefont {Walker},\ and\ \citenamefont {Kondev}}]{Dr16}%
  \BibitemOpen
  \bibfield  {author} {\bibinfo {author} {\bibfnamefont {G.~D.}\ \bibnamefont
  {Dracoulis}}, \bibinfo {author} {\bibfnamefont {P.~M.}\ \bibnamefont
  {Walker}}, \ and\ \bibinfo {author} {\bibfnamefont {F.~G.}\ \bibnamefont
  {Kondev}},\ }\href {\doibase 10.1088/0034-4885/79/7/076301} {\bibfield
  {journal} {\bibinfo  {journal} {Rep. Prog. Phys.}\ }\textbf {\bibinfo
  {volume} {79}},\ \bibinfo {pages} {076301} (\bibinfo {year}
  {2016})}\BibitemShut {NoStop}%
\bibitem [{\citenamefont {Sevrin}\ \emph {et~al.}(1987)\citenamefont {Sevrin},
  \citenamefont {Heyde},\ and\ \citenamefont {Jolie}}]{sevrin1987}%
  \BibitemOpen
  \bibfield  {author} {\bibinfo {author} {\bibfnamefont {A.}~\bibnamefont
  {Sevrin}}, \bibinfo {author} {\bibfnamefont {K.}~\bibnamefont {Heyde}}, \
  and\ \bibinfo {author} {\bibfnamefont {J.}~\bibnamefont {Jolie}},\ }\href
  {\doibase 10.1103/PhysRevC.36.2631} {\bibfield  {journal} {\bibinfo
  {journal} {Phys. Rev. C}\ }\textbf {\bibinfo {volume} {36}},\ \bibinfo
  {pages} {2631} (\bibinfo {year} {1987})}\BibitemShut {NoStop}%
\bibitem [{\citenamefont {Kirch~{\it et al.}}(1995)}]{kirch}%
  \BibitemOpen
  \bibfield  {author} {\bibinfo {author} {\bibfnamefont {K.}~\bibnamefont
  {Kirch~{\it et al.}}},\ }\href {\doibase
  https://doi.org/10.1016/S0375-9474(94)00806-x} {\bibfield  {journal}
  {\bibinfo  {journal} {Nucl. Phys. A}\ }\textbf {\bibinfo {volume} {587}},\
  \bibinfo {pages} {211} (\bibinfo {year} {1995})}\BibitemShut {NoStop}%
\bibitem [{\citenamefont {Caprio}(2011)}]{caprio}%
  \BibitemOpen
  \bibfield  {author} {\bibinfo {author} {\bibfnamefont {M.~A.}\ \bibnamefont
  {Caprio}},\ }\href {\doibase https://doi.org/10.1103/PhysRevC.83.064309}
  {\bibfield  {journal} {\bibinfo  {journal} {Phys. Rev. C}\ }\textbf {\bibinfo
  {volume} {83}},\ \bibinfo {pages} {064309} (\bibinfo {year}
  {2011})}\BibitemShut {NoStop}%
\bibitem [{\citenamefont {Frauendorf}(1993)}]{fra_tilted-cranking}%
  \BibitemOpen
  \bibfield  {author} {\bibinfo {author} {\bibfnamefont {S.}~\bibnamefont
  {Frauendorf}},\ }\href {\doibase 10.1016/0375-9474(93)90546-A} {\bibfield
  {journal} {\bibinfo  {journal} {Nucl. Phys. A}\ }\textbf {\bibinfo {volume}
  {557}},\ \bibinfo {pages} {259} (\bibinfo {year} {1993})}\BibitemShut
  {NoStop}%
\bibitem [{\citenamefont {Frauendorf}(2000)}]{fra-2000}%
  \BibitemOpen
  \bibfield  {author} {\bibinfo {author} {\bibfnamefont {S.}~\bibnamefont
  {Frauendorf}},\ }\href {\doibase 10.1016/0375-9474(00)00308-0} {\bibfield
  {journal} {\bibinfo  {journal} {Nucl. Phys. A}\ }\textbf {\bibinfo {volume}
  {677}},\ \bibinfo {pages} {115} (\bibinfo {year} {2000})}\BibitemShut
  {NoStop}%
\bibitem [{\citenamefont {Walker}\ \emph {et~al.}(1993)\citenamefont {Walker},
  \citenamefont {Yeung}, \citenamefont {Dracoulis}, \citenamefont {Regan},
  \citenamefont {Lane}, \citenamefont {Davidson},\ and\ \citenamefont
  {Stuchbery}}]{180W-walker}%
  \BibitemOpen
  \bibfield  {author} {\bibinfo {author} {\bibfnamefont {P.~M.}\ \bibnamefont
  {Walker}}, \bibinfo {author} {\bibfnamefont {K.~C.}\ \bibnamefont {Yeung}},
  \bibinfo {author} {\bibfnamefont {G.~D.}\ \bibnamefont {Dracoulis}}, \bibinfo
  {author} {\bibfnamefont {P.~H.}\ \bibnamefont {Regan}}, \bibinfo {author}
  {\bibfnamefont {G.~J.}\ \bibnamefont {Lane}}, \bibinfo {author}
  {\bibfnamefont {P.~M.}\ \bibnamefont {Davidson}}, \ and\ \bibinfo {author}
  {\bibfnamefont {A.~E.}\ \bibnamefont {Stuchbery}},\ }\href@noop {} {\bibfield
   {journal} {\bibinfo  {journal} {Phys. Lett. B}\ }\textbf {\bibinfo {volume}
  {309}},\ \bibinfo {pages} {17} (\bibinfo {year} {1993})}\BibitemShut
  {NoStop}%
\bibitem [{\citenamefont {Kutsarova~{\it et al.}}(1995)}]{182os-1995}%
  \BibitemOpen
  \bibfield  {author} {\bibinfo {author} {\bibfnamefont {T.}~\bibnamefont
  {Kutsarova~{\it et al.}}},\ }\href {\doibase 10.1016/0375-9474(94)00819-1}
  {\bibfield  {journal} {\bibinfo  {journal} {Nucl. Phys. A}\ }\textbf
  {\bibinfo {volume} {587}},\ \bibinfo {pages} {111} (\bibinfo {year}
  {1995})}\BibitemShut {NoStop}%
\bibitem [{\citenamefont {Chen}\ \emph {et~al.}(2018)\citenamefont {Chen},
  \citenamefont {Lv}, \citenamefont {Petrache},\ and\ \citenamefont
  {Meng}}]{Ch18}%
  \BibitemOpen
  \bibfield  {author} {\bibinfo {author} {\bibfnamefont {Q.~B.}\ \bibnamefont
  {Chen}}, \bibinfo {author} {\bibfnamefont {B.~F.}\ \bibnamefont {Lv}},
  \bibinfo {author} {\bibfnamefont {C.~M.}\ \bibnamefont {Petrache}}, \ and\
  \bibinfo {author} {\bibfnamefont {J.}~\bibnamefont {Meng}},\ }\href {\doibase
  10.1016/j.physletb.2018.06.030} {\bibfield  {journal} {\bibinfo  {journal}
  {Phys. Lett. B}\ }\textbf {\bibinfo {volume} {782}},\ \bibinfo {pages} {747}
  (\bibinfo {year} {2018})}\BibitemShut {NoStop}%
\bibitem [{\citenamefont {Hara}\ and\ \citenamefont {Sun}(1995)}]{hara-sun}%
  \BibitemOpen
  \bibfield  {author} {\bibinfo {author} {\bibfnamefont {K.}~\bibnamefont
  {Hara}}\ and\ \bibinfo {author} {\bibfnamefont {Y.}~\bibnamefont {Sun}},\
  }\href {\doibase 10.1142/S0218301395000250} {\bibfield  {journal} {\bibinfo
  {journal} {Int. J. Mod. Phys. E}\ }\textbf {\bibinfo {volume} {4}},\ \bibinfo
  {pages} {637} (\bibinfo {year} {1995})}\BibitemShut {NoStop}%
\bibitem [{\citenamefont {Sun}\ \emph {et~al.}(2001)\citenamefont {Sun},
  \citenamefont {Zhang},\ and\ \citenamefont {Guidry}}]{Sun-psm}%
  \BibitemOpen
  \bibfield  {author} {\bibinfo {author} {\bibfnamefont {Y.}~\bibnamefont
  {Sun}}, \bibinfo {author} {\bibfnamefont {J.-Y.}\ \bibnamefont {Zhang}}, \
  and\ \bibinfo {author} {\bibfnamefont {M.}~\bibnamefont {Guidry}},\ }\href
  {\doibase https://doi.org/10.1103/PhysRevC.63.047306} {\bibfield  {journal}
  {\bibinfo  {journal} {Phys. Rev. C}\ }\textbf {\bibinfo {volume} {63}},\
  \bibinfo {pages} {047306} (\bibinfo {year} {2001})}\BibitemShut {NoStop}%
\bibitem [{\citenamefont {Valiente-Dob\'on~{\it et al.}}(2014)}]{val2014}%
  \BibitemOpen
  \bibfield  {author} {\bibinfo {author} {\bibfnamefont {J.~J.}\ \bibnamefont
  {Valiente-Dob\'on~{\it et al.}}},\ }\href@noop {} {\bibfield  {journal}
  {\bibinfo  {journal} {INFN LNL Annual Report}\ } (\bibinfo {year}
  {2014})}\BibitemShut {NoStop}%
\bibitem [{\citenamefont {Testov~{\it et al.}}(2018)}]{testov2018}%
  \BibitemOpen
  \bibfield  {author} {\bibinfo {author} {\bibfnamefont {D.}~\bibnamefont
  {Testov~{\it et al.}}},\ }\href@noop {} {\bibfield  {journal} {\bibinfo
  {journal} {arXiv:1903.01296 [nucl-ex]}\ } (\bibinfo {year}
  {2018})}\BibitemShut {NoStop}%
\bibitem [{\citenamefont {Testov}(2019)}]{EUCLIDES}%
  \BibitemOpen
  \bibfield  {author} {\bibinfo {author} {\bibfnamefont {D.}~\bibnamefont
  {Testov}},\ }\href {\doibase https://doi.org/10.1140/epja/i2019-12714-6}
  {\bibfield  {journal} {\bibinfo  {journal} {Eur. Phys. J. A}\ }\textbf
  {\bibinfo {volume} {55}},\ \bibinfo {pages} {47} (\bibinfo {year}
  {2019})}\BibitemShut {NoStop}%
\bibitem [{\citenamefont {Skeppstedt}\ \emph {et~al.}(1999)\citenamefont
  {Skeppstedt}, \citenamefont {Roth}, \citenamefont {Lindstr�m},\ and\
  \citenamefont {{\it et al.}}}]{nw1}%
  \BibitemOpen
  \bibfield  {author} {\bibinfo {author} {\bibfnamefont {O.}~\bibnamefont
  {Skeppstedt}}, \bibinfo {author} {\bibfnamefont {H.~A.}\ \bibnamefont
  {Roth}}, \bibinfo {author} {\bibfnamefont {L.}~\bibnamefont {Lindstr�m}}, \
  and\ \bibinfo {author} {\bibnamefont {{\it et al.}}},\ }\href {\doibase
  10/1016/S0168-9002(98)01208-X} {\bibfield  {journal} {\bibinfo  {journal}
  {Nucl. Instrum. Meth. Phys. Res. A}\ }\textbf {\bibinfo {volume} {421}},\
  \bibinfo {pages} {531} (\bibinfo {year} {1999})}\BibitemShut {NoStop}%
\bibitem [{\citenamefont {Ljungvall}\ \emph {et~al.}(2004)\citenamefont
  {Ljungvall}, \citenamefont {Palacz},\ and\ \citenamefont {Nyberg}}]{nw2}%
  \BibitemOpen
  \bibfield  {author} {\bibinfo {author} {\bibfnamefont {J.}~\bibnamefont
  {Ljungvall}}, \bibinfo {author} {\bibfnamefont {M.}~\bibnamefont {Palacz}}, \
  and\ \bibinfo {author} {\bibfnamefont {J.}~\bibnamefont {Nyberg}},\ }\href
  {\doibase 10.1016/j.nima.2004.05.032} {\bibfield  {journal} {\bibinfo
  {journal} {Nucl. Instrum. Meth. Phys. Res. A}\ }\textbf {\bibinfo {volume}
  {528}},\ \bibinfo {pages} {741} (\bibinfo {year} {2004})}\BibitemShut
  {NoStop}%
\bibitem [{\citenamefont {Berti}\ \emph {et~al.}(2014)\citenamefont {Berti},
  \citenamefont {Biasotto}, \citenamefont {Fantinel}, \citenamefont
  {Gozzelino}, \citenamefont {Gulmini},\ and\ \citenamefont
  {Toniolo}}]{Berti2015}%
  \BibitemOpen
  \bibfield  {author} {\bibinfo {author} {\bibfnamefont {L.}~\bibnamefont
  {Berti}}, \bibinfo {author} {\bibfnamefont {M.}~\bibnamefont {Biasotto}},
  \bibinfo {author} {\bibfnamefont {S.}~\bibnamefont {Fantinel}}, \bibinfo
  {author} {\bibfnamefont {A.}~\bibnamefont {Gozzelino}}, \bibinfo {author}
  {\bibfnamefont {M.}~\bibnamefont {Gulmini}}, \ and\ \bibinfo {author}
  {\bibfnamefont {N.}~\bibnamefont {Toniolo}},\ }\href@noop {} {\bibfield
  {journal} {\bibinfo  {journal} {LNL INFN Annual Report}\ ,\ \bibinfo {pages}
  {93}} (\bibinfo {year} {2014})}\BibitemShut {NoStop}%
\bibitem [{\citenamefont {Guo~{\it et al.}}()}]{130ba-gs}%
  \BibitemOpen
  \bibfield  {author} {\bibinfo {author} {\bibfnamefont {S.}~\bibnamefont
  {Guo~{\it et al.}}},\ }\href@noop {} {\bibinfo  {journal} {{\it to be
  published}}\ }\BibitemShut {NoStop}%
\bibitem [{SM()}]{SM}%
  \BibitemOpen
\bibfield  {journal} {  }\href@noop {} {\bibinfo  {journal} {See Supplemental
  Material at [URL will be inserted by publisher]}\ }\BibitemShut {NoStop}%
\bibitem [{\citenamefont {Kaur~{\it et al}}(2014)}]{Ka14}%
  \BibitemOpen
\bibfield  {journal} {  }\bibfield  {author} {\bibinfo {author} {\bibfnamefont
  {N.}~\bibnamefont {Kaur~{\it et al}}},\ }\href {\doibase
  10.1140/epja/i2014-14005-2} {\bibfield  {journal} {\bibinfo  {journal} {Eur.
  Phys. J. A}\ }\textbf {\bibinfo {volume} {50}},\ \bibinfo {pages} {5}
  (\bibinfo {year} {2014})}\BibitemShut {NoStop}%
\bibitem [{\citenamefont {Bengtsson}\ and\ \citenamefont
  {Frauendorf}(1979)}]{bf1979}%
  \BibitemOpen
  \bibfield  {author} {\bibinfo {author} {\bibfnamefont {R.}~\bibnamefont
  {Bengtsson}}\ and\ \bibinfo {author} {\bibfnamefont {S.}~\bibnamefont
  {Frauendorf}},\ }\href {\doibase 10.1016/0375-9474(79)90322-1} {\bibfield
  {journal} {\bibinfo  {journal} {Nucl. Phys. A}\ }\textbf {\bibinfo {volume}
  {327}},\ \bibinfo {pages} {139} (\bibinfo {year} {1979})}\BibitemShut
  {NoStop}%
\bibitem [{\citenamefont {Bengtsson}\ \emph {et~al.}(1986)\citenamefont
  {Bengtsson}, \citenamefont {Frauendorf},\ and\ \citenamefont
  {May}}]{bengtsson1986}%
  \BibitemOpen
  \bibfield  {author} {\bibinfo {author} {\bibfnamefont {R.}~\bibnamefont
  {Bengtsson}}, \bibinfo {author} {\bibfnamefont {S.}~\bibnamefont
  {Frauendorf}}, \ and\ \bibinfo {author} {\bibfnamefont {F.~R.}\ \bibnamefont
  {May}},\ }\href {\doibase 10.1016/0092-640X(86)90028-8} {\bibfield  {journal}
  {\bibinfo  {journal} {Atomic Datas and Nuclear Data Tables}\ }\textbf
  {\bibinfo {volume} {35}},\ \bibinfo {pages} {15} (\bibinfo {year}
  {1986})}\BibitemShut {NoStop}%
\bibitem [{\citenamefont {Orce~{\it et al.}}(2006)}]{128xe-orce}%
  \BibitemOpen
  \bibfield  {author} {\bibinfo {author} {\bibfnamefont {J.~N.}\ \bibnamefont
  {Orce~{\it et al.}}},\ }\href {\doibase 10.1103/PhysRevC.74.034318}
  {\bibfield  {journal} {\bibinfo  {journal} {Phys. Rev. C}\ }\textbf {\bibinfo
  {volume} {74}},\ \bibinfo {pages} {034318} (\bibinfo {year}
  {2006})}\BibitemShut {NoStop}%
\bibitem [{\citenamefont {Frauendorf}(1991)}]{126ba-ward}%
  \BibitemOpen
  \bibfield  {author} {\bibinfo {author} {\bibfnamefont {S.}~\bibnamefont
  {Frauendorf}},\ }\href {\doibase 10.1016/0375-9474(91)90798-B} {\bibfield
  {journal} {\bibinfo  {journal} {Nucl. Phys. A}\ }\textbf {\bibinfo {volume}
  {529}},\ \bibinfo {pages} {315} (\bibinfo {year} {1991})}\BibitemShut
  {NoStop}%
\bibitem [{\citenamefont {Vogel~{\it et al}}(1999)}]{128ba}%
  \BibitemOpen
  \bibfield  {author} {\bibinfo {author} {\bibfnamefont {O.}~\bibnamefont
  {Vogel~{\it et al}}},\ }\href {\doibase 10.1007/s100500050238} {\bibfield
  {journal} {\bibinfo  {journal} {Eur. Phys. J. A}\ }\textbf {\bibinfo {volume}
  {4}},\ \bibinfo {pages} {323} (\bibinfo {year} {1999})}\BibitemShut {NoStop}%
\bibitem [{\citenamefont {Juutinen~{\it et al.}}(1995)}]{132ba}%
  \BibitemOpen
  \bibfield  {author} {\bibinfo {author} {\bibfnamefont {S.}~\bibnamefont
  {Juutinen~{\it et al.}}},\ }\href {\doibase 10.1103/PhysRevC.52.2946}
  {\bibfield  {journal} {\bibinfo  {journal} {Phys. Rev. C}\ }\textbf {\bibinfo
  {volume} {52}},\ \bibinfo {pages} {2946} (\bibinfo {year}
  {1995})}\BibitemShut {NoStop}%
\bibitem [{\citenamefont {Petrache~{\it et al.}}(2016)}]{134ce-petrache}%
  \BibitemOpen
  \bibfield  {author} {\bibinfo {author} {\bibfnamefont {C.~M.}\ \bibnamefont
  {Petrache~{\it et al.}}},\ }\href {\doibase 10.1103/PhysRevC.93.064305}
  {\bibfield  {journal} {\bibinfo  {journal} {Phys. Rev. C}\ }\textbf {\bibinfo
  {volume} {93}},\ \bibinfo {pages} {064305} (\bibinfo {year}
  {2016})}\BibitemShut {NoStop}%
\bibitem [{\citenamefont {Lv~{\it et al.}}(2018)}]{136nd-lv}%
  \BibitemOpen
  \bibfield  {author} {\bibinfo {author} {\bibfnamefont {B.~F.}\ \bibnamefont
  {Lv~{\it et al.}}},\ }\href {\doibase 10.1103/PhysRevC.98.044304} {\bibfield
  {journal} {\bibinfo  {journal} {Phys. Rev. C}\ }\textbf {\bibinfo {volume}
  {98}},\ \bibinfo {pages} {044304} (\bibinfo {year} {2018})}\BibitemShut
  {NoStop}%
\bibitem [{\citenamefont {Petrache~{\it et al.}}(2012)}]{138nd-petrache}%
  \BibitemOpen
  \bibfield  {author} {\bibinfo {author} {\bibfnamefont {C.~M.}\ \bibnamefont
  {Petrache~{\it et al.}}},\ }\href {\doibase 10.1103/PhysRevC.86.044321}
  {\bibfield  {journal} {\bibinfo  {journal} {Phys. Rev. C}\ }\textbf {\bibinfo
  {volume} {86}},\ \bibinfo {pages} {044321} (\bibinfo {year}
  {2012})}\BibitemShut {NoStop}%
\bibitem [{\citenamefont {Paul~{\it et al.}}(1997)}]{132ce-paul}%
  \BibitemOpen
  \bibfield  {author} {\bibinfo {author} {\bibfnamefont {E.~S.}\ \bibnamefont
  {Paul~{\it et al.}}},\ }\href@noop {} {\bibfield  {journal} {\bibinfo
  {journal} {Nucl. Phys. A}\ }\textbf {\bibinfo {volume} {619}},\ \bibinfo
  {pages} {177} (\bibinfo {year} {1997})}\BibitemShut {NoStop}%
\bibitem [{\citenamefont {Byrne~{\it et al.}}(1992)}]{129ba-byrne}%
  \BibitemOpen
  \bibfield  {author} {\bibinfo {author} {\bibfnamefont {A.~P.}\ \bibnamefont
  {Byrne~{\it et al.}}},\ }\href@noop {} {\bibfield  {journal} {\bibinfo
  {journal} {Nucl. Phys. A}\ }\textbf {\bibinfo {volume} {548}},\ \bibinfo
  {pages} {131} (\bibinfo {year} {1992})}\BibitemShut {NoStop}%
\bibitem [{\citenamefont {Chakraborty~{\it et al.}}(2006)}]{127xe-Chakraborty}%
  \BibitemOpen
  \bibfield  {author} {\bibinfo {author} {\bibfnamefont {S.}~\bibnamefont
  {Chakraborty~{\it et al.}}},\ }\href {\doibase 10.1103/PhysRevC.74.034318}
  {\bibfield  {journal} {\bibinfo  {journal} {Phys. Rev. C}\ }\textbf {\bibinfo
  {volume} {74}},\ \bibinfo {pages} {034318} (\bibinfo {year}
  {2006})}\BibitemShut {NoStop}%
\bibitem [{\citenamefont {Walker}\ and\ \citenamefont
  {Xu}(2016)}]{Walker2016Phys.Scr013010}%
  \BibitemOpen
  \bibfield  {author} {\bibinfo {author} {\bibfnamefont {P.~M.}\ \bibnamefont
  {Walker}}\ and\ \bibinfo {author} {\bibfnamefont {F.~R.}\ \bibnamefont
  {Xu}},\ }\href {\doibase 10.1088/0031-8949/91/1/013010} {\bibfield  {journal}
  {\bibinfo  {journal} {Phys. Scr.}\ }\textbf {\bibinfo {volume} {91}},\
  \bibinfo {pages} {013010} (\bibinfo {year} {2016})}\BibitemShut {NoStop}%
\bibitem [{\citenamefont {Sun~{\it et al.}}(1983)}]{130ba-sun}%
  \BibitemOpen
  \bibfield  {author} {\bibinfo {author} {\bibfnamefont {X.}~\bibnamefont
  {Sun~{\it et al.}}},\ }\href {\doibase 10.1103/PhysRevC.28.1167} {\bibfield
  {journal} {\bibinfo  {journal} {Phys. Rev. C}\ }\textbf {\bibinfo {volume}
  {28}},\ \bibinfo {pages} {1167} (\bibinfo {year} {1983})}\BibitemShut
  {NoStop}%
\bibitem [{\citenamefont {Rohozinski}\ \emph {et~al.}(1977)\citenamefont
  {Rohozinski}, \citenamefont {Dobaczewski}, \citenamefont {Nerlo-Pomorska},
  \citenamefont {Pomorski},\ and\ \citenamefont {Srebrny}}]{rohozinski1977}%
  \BibitemOpen
  \bibfield  {author} {\bibinfo {author} {\bibfnamefont {S.~G.}\ \bibnamefont
  {Rohozinski}}, \bibinfo {author} {\bibfnamefont {J.}~\bibnamefont
  {Dobaczewski}}, \bibinfo {author} {\bibfnamefont {B.}~\bibnamefont
  {Nerlo-Pomorska}}, \bibinfo {author} {\bibfnamefont {K.}~\bibnamefont
  {Pomorski}}, \ and\ \bibinfo {author} {\bibfnamefont {J.}~\bibnamefont
  {Srebrny}},\ }\href {\doibase 10.1016/0375-9474(77)90358-X} {\bibfield
  {journal} {\bibinfo  {journal} {Nucl. Phys. A}\ }\textbf {\bibinfo {volume}
  {292}},\ \bibinfo {pages} {66} (\bibinfo {year} {1977})}\BibitemShut
  {NoStop}%
\bibitem [{\citenamefont {Faessler}\ \emph {et~al.}(1985)\citenamefont
  {Faessler}, \citenamefont {Kuyucak}, \citenamefont {Petrovici},\ and\
  \citenamefont {Petersen}}]{faessler1985}%
  \BibitemOpen
  \bibfield  {author} {\bibinfo {author} {\bibfnamefont {A.}~\bibnamefont
  {Faessler}}, \bibinfo {author} {\bibfnamefont {S.}~\bibnamefont {Kuyucak}},
  \bibinfo {author} {\bibfnamefont {A.}~\bibnamefont {Petrovici}}, \ and\
  \bibinfo {author} {\bibfnamefont {L.}~\bibnamefont {Petersen}},\ }\href
  {\doibase 10.1016/0375-9474(85)90119-8} {\bibfield  {journal} {\bibinfo
  {journal} {Nucl. Phys. A}\ }\textbf {\bibinfo {volume} {438}},\ \bibinfo
  {pages} {78} (\bibinfo {year} {1985})}\BibitemShut {NoStop}%
\bibitem [{\citenamefont {Hammaren}\ \emph {et~al.}(1986)\citenamefont
  {Hammaren}, \citenamefont {Schmid}, \citenamefont {Gr\"ummer}, \citenamefont
  {Faessler},\ and\ \citenamefont {Fladt}}]{hammaren1986}%
  \BibitemOpen
  \bibfield  {author} {\bibinfo {author} {\bibfnamefont {E.}~\bibnamefont
  {Hammaren}}, \bibinfo {author} {\bibfnamefont {K.~W.}\ \bibnamefont
  {Schmid}}, \bibinfo {author} {\bibfnamefont {F.}~\bibnamefont {Gr\"ummer}},
  \bibinfo {author} {\bibfnamefont {A.}~\bibnamefont {Faessler}}, \ and\
  \bibinfo {author} {\bibfnamefont {B.}~\bibnamefont {Fladt}},\ }\href
  {\doibase 10.1016/0375-9474(86)90271-X} {\bibfield  {journal} {\bibinfo
  {journal} {Nucl. Phys. A}\ }\textbf {\bibinfo {volume} {454}},\ \bibinfo
  {pages} {301} (\bibinfo {year} {1986})}\BibitemShut {NoStop}%
\bibitem [{\citenamefont {Wyss}\ \emph
  {et~al.}(1988{\natexlab{b}})\citenamefont {Wyss}, \citenamefont {Johnson},
  \citenamefont {Bengtsson},\ and\ \citenamefont {Nazarewicz}}]{wyss1988}%
  \BibitemOpen
  \bibfield  {author} {\bibinfo {author} {\bibfnamefont {R.~A.}\ \bibnamefont
  {Wyss}}, \bibinfo {author} {\bibfnamefont {A.}~\bibnamefont {Johnson}},
  \bibinfo {author} {\bibfnamefont {R.}~\bibnamefont {Bengtsson}}, \ and\
  \bibinfo {author} {\bibfnamefont {W.}~\bibnamefont {Nazarewicz}},\ }\href
  {\doibase 10.1007/BF01283782} {\bibfield  {journal} {\bibinfo  {journal} {Z.
  Phys. A}\ }\textbf {\bibinfo {volume} {329}},\ \bibinfo {pages} {255}
  (\bibinfo {year} {1988}{\natexlab{b}})}\BibitemShut {NoStop}%
\bibitem [{\citenamefont {Wyss~{\it et al.}}(1989)}]{wyss1989}%
  \BibitemOpen
  \bibfield  {author} {\bibinfo {author} {\bibfnamefont {R.~A.}\ \bibnamefont
  {Wyss~{\it et al.}}},\ }\href {\doibase 10.1016/0375-9474(89)90378-3}
  {\bibfield  {journal} {\bibinfo  {journal} {Nucl. Phys. A}\ }\textbf
  {\bibinfo {volume} {337}},\ \bibinfo {pages} {505} (\bibinfo {year}
  {1989})}\BibitemShut {NoStop}%
\bibitem [{\citenamefont {Granderath}\ and\ \citenamefont
  {Mantica}(1996)}]{granderath1996}%
  \BibitemOpen
  \bibfield  {author} {\bibinfo {author} {\bibfnamefont {A.}~\bibnamefont
  {Granderath}}\ and\ \bibinfo {author} {\bibfnamefont {R.~A.}\ \bibnamefont
  {Mantica}, \bibfnamefont {P.~F. Wyss~{\it et al.}}},\ }\href {\doibase
  10.1016/0375-9474(95)00484-X} {\bibfield  {journal} {\bibinfo  {journal}
  {Nucl. Phys. A}\ }\textbf {\bibinfo {volume} {427}},\ \bibinfo {pages} {505}
  (\bibinfo {year} {1996})}\BibitemShut {NoStop}%
\bibitem [{\citenamefont {Meng~{(Ed.)}}()}]{meng2016}%
  \BibitemOpen
  \bibfield  {author} {\bibinfo {author} {\bibfnamefont {J.}~\bibnamefont
  {Meng~{(Ed.)}}},\ }\href@noop {} {\bibinfo  {journal} {{\it Relativistic
  Density Functional for Nuclear Structure}, International Review of Nuclear
  Physics, vol.10, World Scientific, Singapore, 2016}\ }\BibitemShut {NoStop}%
\bibitem [{\citenamefont {Zhao}\ \emph {et~al.}(2010)\citenamefont {Zhao},
  \citenamefont {Li}, \citenamefont {Yao},\ and\ \citenamefont
  {Meng}}]{Zhao2010Phys.Rev.C054319}%
  \BibitemOpen
\bibfield  {journal} {  }\bibfield  {author} {\bibinfo {author} {\bibfnamefont
  {P.~W.}\ \bibnamefont {Zhao}}, \bibinfo {author} {\bibfnamefont {Z.~P.}\
  \bibnamefont {Li}}, \bibinfo {author} {\bibfnamefont {J.~M.}\ \bibnamefont
  {Yao}}, \ and\ \bibinfo {author} {\bibfnamefont {J.}~\bibnamefont {Meng}},\
  }\href {\doibase 10.1103/PhysRevC.82.054319} {\bibfield  {journal} {\bibinfo
  {journal} {Phys. Rev. C}\ }\textbf {\bibinfo {volume} {82}},\ \bibinfo
  {pages} {054319} (\bibinfo {year} {2010})}\BibitemShut {NoStop}%
\bibitem [{\citenamefont {Qi}\ \emph {et~al.}(2009)\citenamefont {Qi},
  \citenamefont {Zhang}, \citenamefont {Meng}, \citenamefont {Wang},\ and\
  \citenamefont {Frauendorf}}]{135nd-prm}%
  \BibitemOpen
  \bibfield  {author} {\bibinfo {author} {\bibfnamefont {B.}~\bibnamefont
  {Qi}}, \bibinfo {author} {\bibfnamefont {S.~Q.}\ \bibnamefont {Zhang}},
  \bibinfo {author} {\bibfnamefont {J.}~\bibnamefont {Meng}}, \bibinfo {author}
  {\bibfnamefont {S.~Y.}\ \bibnamefont {Wang}}, \ and\ \bibinfo {author}
  {\bibfnamefont {S.}~\bibnamefont {Frauendorf}},\ }\href {\doibase
  10.1016j.physletb.2009.02.061} {\bibfield  {journal} {\bibinfo  {journal}
  {Phys. Lett. B}\ }\textbf {\bibinfo {volume} {675}},\ \bibinfo {pages} {175}
  (\bibinfo {year} {2009})}\BibitemShut {NoStop}%
\bibitem [{\citenamefont {Streck}\ \emph {et~al.}(2018)\citenamefont {Streck},
  \citenamefont {Chen}, \citenamefont {Kaiser},\ and\ \citenamefont
  {Meissner}}]{streck2018}%
  \BibitemOpen
  \bibfield  {author} {\bibinfo {author} {\bibfnamefont {E.}~\bibnamefont
  {Streck}}, \bibinfo {author} {\bibfnamefont {Q.~B.}\ \bibnamefont {Chen}},
  \bibinfo {author} {\bibfnamefont {N.}~\bibnamefont {Kaiser}}, \ and\ \bibinfo
  {author} {\bibfnamefont {U.-G.}\ \bibnamefont {Meissner}},\ }\href {\doibase
  10.1103/PhysRevC.98.044314} {\bibfield  {journal} {\bibinfo  {journal} {Phys.
  Rev. C}\ }\textbf {\bibinfo {volume} {98}},\ \bibinfo {pages} {044314}
  (\bibinfo {year} {2018})}\BibitemShut {NoStop}%
\bibitem [{\citenamefont {Bohr}\ and\ \citenamefont {Mottelson}()}]{BM}%
  \BibitemOpen
  \bibfield  {author} {\bibinfo {author} {\bibfnamefont {A.}~\bibnamefont
  {Bohr}}\ and\ \bibinfo {author} {\bibfnamefont {B.~R.}\ \bibnamefont
  {Mottelson}},\ }\href@noop {} {\bibinfo  {journal} {{\it Nuclear Structure},
  Vol. I (Benjamin, New York), (1975)}\ }\BibitemShut {NoStop}%
\bibitem [{\citenamefont {Sun}(2016)}]{Sun-2016}%
  \BibitemOpen
\bibfield  {journal} {  }\bibfield  {author} {\bibinfo {author} {\bibfnamefont
  {Y.}~\bibnamefont {Sun}},\ }\href@noop {} {\bibfield  {journal} {\bibinfo
  {journal} {Phys. Scr.}\ }\textbf {\bibinfo {volume} {91}},\ \bibinfo {pages}
  {043005} (\bibinfo {year} {2016})}\BibitemShut {NoStop}%
\bibitem [{\citenamefont {Sheikh}\ \emph {et~al.}(1998)\citenamefont {Sheikh},
  \citenamefont {Sun},\ and\ \citenamefont {Walker}}]{Sheikh1998Phys.Rev.R26}%
  \BibitemOpen
  \bibfield  {author} {\bibinfo {author} {\bibfnamefont {J.~A.}\ \bibnamefont
  {Sheikh}}, \bibinfo {author} {\bibfnamefont {Y.}~\bibnamefont {Sun}}, \ and\
  \bibinfo {author} {\bibfnamefont {P.~M.}\ \bibnamefont {Walker}},\ }\href
  {\doibase 10.1103/PhysRevC.57.} {\bibfield  {journal} {\bibinfo  {journal}
  {Phys. Rev. C}\ }\textbf {\bibinfo {volume} {57}},\ \bibinfo {pages} {R26}
  (\bibinfo {year} {1998})}\BibitemShut {NoStop}%
\bibitem [{\citenamefont {Wang~{\it et al.}}(2014)}]{LJ-2014}%
  \BibitemOpen
  \bibfield  {author} {\bibinfo {author} {\bibfnamefont {L.-J.}\ \bibnamefont
  {Wang~{\it et al.}}},\ }\href {\doibase
  https://doi.org/10.1103/PhysRevC.90.011303} {\bibfield  {journal} {\bibinfo
  {journal} {Phys. Rev. C}\ }\textbf {\bibinfo {volume} {90}},\ \bibinfo
  {pages} {011303} (\bibinfo {year} {2014})}\BibitemShut {NoStop}%
\bibitem [{\citenamefont {Liu~{\it et al.}}(2011)}]{YX-2011}%
  \BibitemOpen
  \bibfield  {author} {\bibinfo {author} {\bibfnamefont {Y.-X.}\ \bibnamefont
  {Liu~{\it et al.}}},\ }\href {\doibase 10.1016/j.nuclphysa.2011.03.010}
  {\bibfield  {journal} {\bibinfo  {journal} {Nucl. Phys. A}\ }\textbf
  {\bibinfo {volume} {858}},\ \bibinfo {pages} {11} (\bibinfo {year}
  {2011})}\BibitemShut {NoStop}%
\bibitem [{\citenamefont {Petrache~{\it et al.}}(1996)}]{costel-1996}%
  \BibitemOpen
  \bibfield  {author} {\bibinfo {author} {\bibfnamefont {C.~M.}\ \bibnamefont
  {Petrache~{\it et al.}}},\ }\href {\doibase
  https://doi.org/10.1103/PhysRevC.53.R2581} {\bibfield  {journal} {\bibinfo
  {journal} {Phys. Rev. C}\ }\textbf {\bibinfo {volume} {53}},\ \bibinfo
  {pages} {R2581} (\bibinfo {year} {1996})}\BibitemShut {NoStop}%
\bibitem [{\citenamefont {Petrache~{\it et al.}}(1997)}]{costel-1997}%
  \BibitemOpen
  \bibfield  {author} {\bibinfo {author} {\bibfnamefont {C.~M.}\ \bibnamefont
  {Petrache~{\it et al.}}},\ }\href {\doibase
  https://doi.org/10.1016/S0375-9474(97)00036-5} {\bibfield  {journal}
  {\bibinfo  {journal} {Nucl. Phys. A}\ }\textbf {\bibinfo {volume} {617}},\
  \bibinfo {pages} {249} (\bibinfo {year} {1997})}\BibitemShut {NoStop}%
\bibitem [{\citenamefont {M\"oller}\ \emph {et~al.}(2016)\citenamefont
  {M\"oller}, \citenamefont {Sierk}, \citenamefont {Ichikawa},\ and\
  \citenamefont {Sagawa}}]{moller2016}%
  \BibitemOpen
  \bibfield  {author} {\bibinfo {author} {\bibfnamefont {P.}~\bibnamefont
  {M\"oller}}, \bibinfo {author} {\bibfnamefont {A.~J.}\ \bibnamefont {Sierk}},
  \bibinfo {author} {\bibfnamefont {T.}~\bibnamefont {Ichikawa}}, \ and\
  \bibinfo {author} {\bibfnamefont {H.}~\bibnamefont {Sagawa}},\ }\href
  {\doibase 10.1016/j.adt.2015.10.002} {\bibfield  {journal} {\bibinfo
  {journal} {Atomic Datas and Nuclear Data Tables}\ }\textbf {\bibinfo {volume}
  {109-110}},\ \bibinfo {pages} {1} (\bibinfo {year} {2016})}\BibitemShut
  {NoStop}%
\bibitem [{\citenamefont {Sheikh}\ and\ \citenamefont
  {Hara}(1999)}]{sheikh-TPSM}%
  \BibitemOpen
  \bibfield  {author} {\bibinfo {author} {\bibfnamefont {J.~A.}\ \bibnamefont
  {Sheikh}}\ and\ \bibinfo {author} {\bibfnamefont {K.}~\bibnamefont {Hara}},\
  }\href {\doibase https://doi.org/10.1103/PhysRevLett.82.3968} {\bibfield
  {journal} {\bibinfo  {journal} {Phys. Rev. Lett.}\ }\textbf {\bibinfo
  {volume} {82}},\ \bibinfo {pages} {3968} (\bibinfo {year}
  {1999})}\BibitemShut {NoStop}%
\end{thebibliography}%
\end{document}


\title{Diversity of shapes and rotations in the $\gamma$-soft $^{130}$Ba nucleus: first observation of a $t$-band in the A=130 ma ss region}

\begin{table}
\caption{Branching ratios, experimental ratios of reduced transition probabilities for band D1 and the $t$-band. } \label{tab1}
\begin{tabular}{llll}
\hline\hline
 {$I^{\pi}_{i}$ } & {$\lambda$ }& {$B(M1)/B(E2)$}&{$B(E2)/B(E2)$}\\
\hline
$t$-band &  &  &       \\
$10^+$& 0.8(2)   & 2.2(6)   &0.19$^{+19}_{-14}$\\
$11^+$& $\geq$32   & $\leq$0.4   &\\
$12^+$&0.21(12)   &3.7(21)   &4.0$^{+27}_{-26}$\\
$13^+$& $\geq$2.1   & $\leq$2.8   &\\
D1 &    &    &\\
$14^+$& $\leq$0.16  & $\geq$8.1  &$\geq$3.6\\
$15^+$& $\leq$0.19   & $\geq$8.1   &$\geq$1.4\\
$16^+$& $\leq$0.09   & $\geq$21   &$\geq$5.1\\
$17^+$& $\leq$0.12   & $\geq$20  &$\geq$5.1\\
$18^+$&  $\leq$0.5  &$\geq$11    &\\
$19^+$& 1.1(7) &3.3(21  &\\
\hline\hline
\end{tabular}
\end{table}

\begin{table}
\begin{center}
\caption{ $J^{(2)} = {{\Delta I} \over {\Delta \omega}}$, $i_x(\omega=0)$, deformation, configuration and coupling of angular momentum values for selected bands of $^{130}$Ba.} \label{tab1}
\begin{tabular}{llll}
\hline\hline
 Band & $J^{(2)}$ & $i_x(\omega=0)$&Comments\\
\hline
S1-even   & 24 & 6 &  prolate, $\pi h^2$, RAL\\
S2o, S2o'     & 15 & 7 &  oblate, $\nu h^{2}$, RAL\\
K=$8^-$high & 20 &  10  &  prolate, $ \nu h^{-1}g^{-1} \otimes \pi h^2$, DAL\\
D1            & 22 & 8 &  prolate, $\nu h^2 \otimes \pi h^{-2}$, FAL\\
\hline
S1-middle    & 36 & 0 & prolate, $\pi h^2 \otimes \nu h^{2}$, RAL\\
S1-high    & 36 & 0 & prolate, $\pi h^2 \otimes \nu h^{4}$, RAL\\
S1'-odd    & 30 & 4 & prolate, $\pi h^2$, $\gamma$-vibration/wobbling \\
S2p & 30 & 1 &  prolate, $\nu h^{2}$, FAL\\
S2o-high    & 24 & 8 &  oblate,  $\nu h^{4}$, RAL\\
\hline
K=$8^-$low & 20 &  6 &  prolate, $\nu h^{-1}g^{-1}$, DAL\\
t-band      & 20 & 6 &  prolate, $\nu h^{-2}$, FAL\\
\hline\hline
\end{tabular}
\end{center}
\end{table}

\begin{figure*}[]
\includegraphics[angle=0,width=16.5cm]{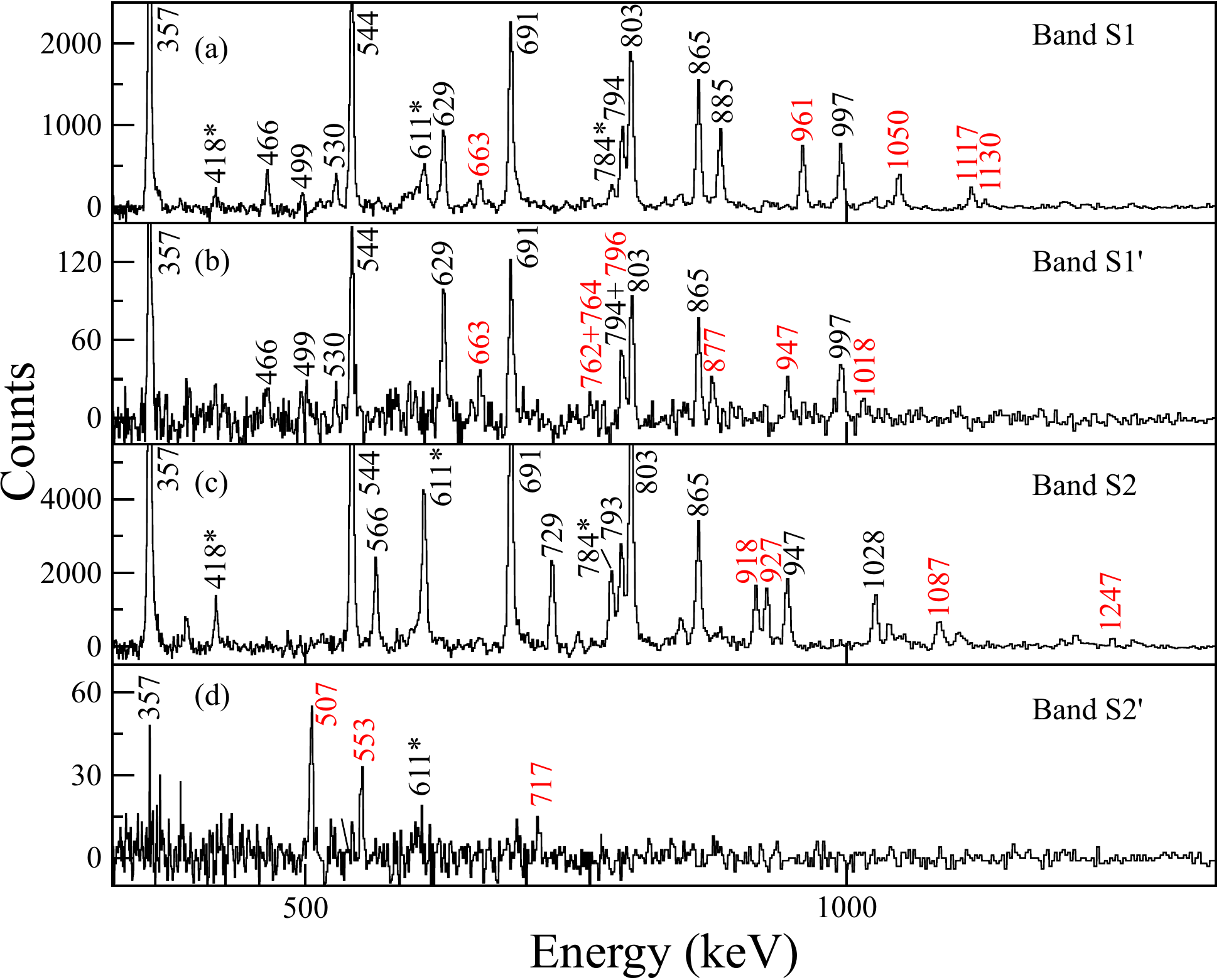}
\caption{(Color online) Spectra for bands S1, S1', S2 and S2' obtained by double gating on the in-band transitions. Transitions belonging to the new structure are marked in red, while those marked with asterisks are identified contaminants.}
\label{fig8}
\end{figure*}

\begin{figure*}[]
\includegraphics[angle=0,width=8.5cm]{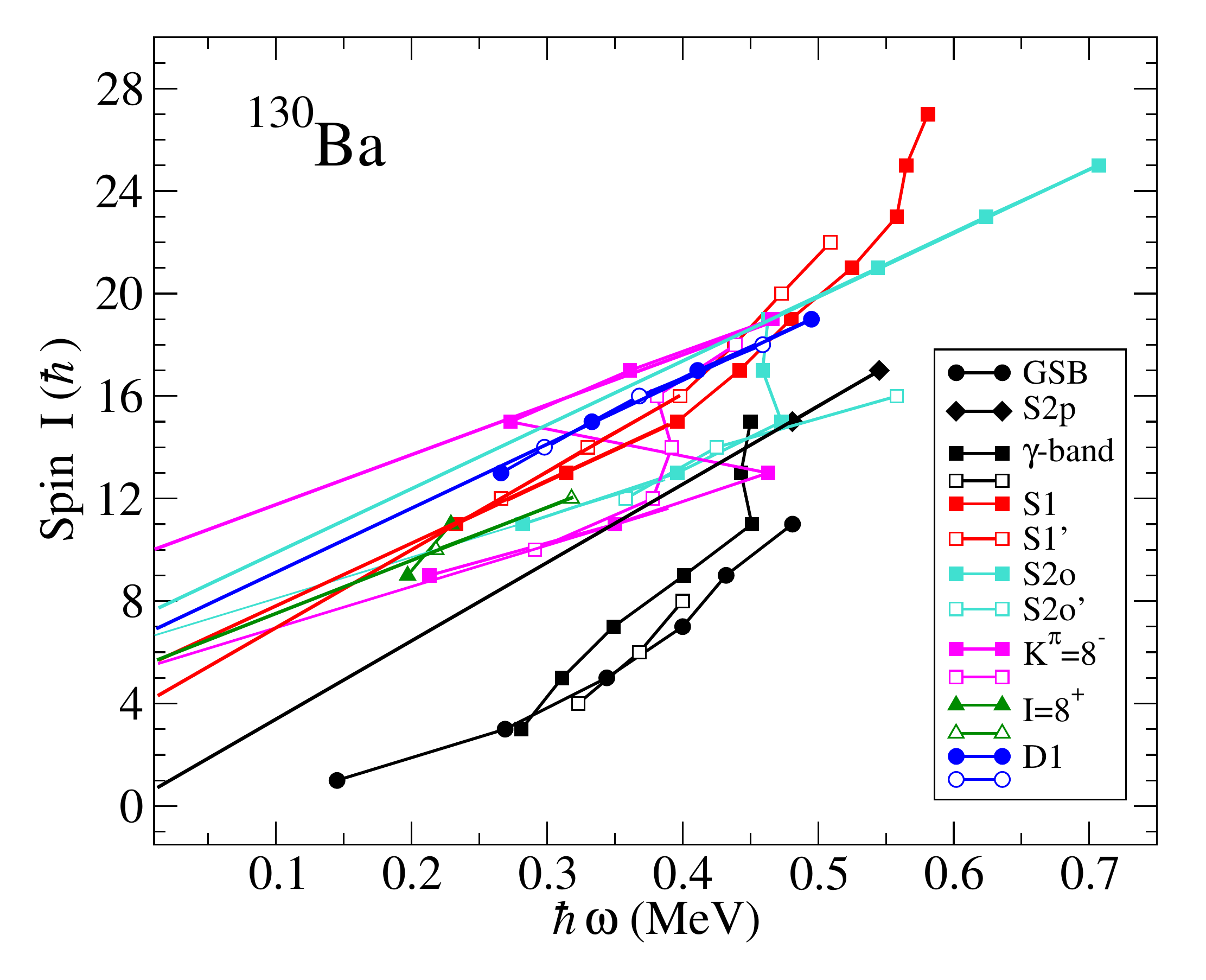}
\includegraphics[angle=0,width=8.5cm]{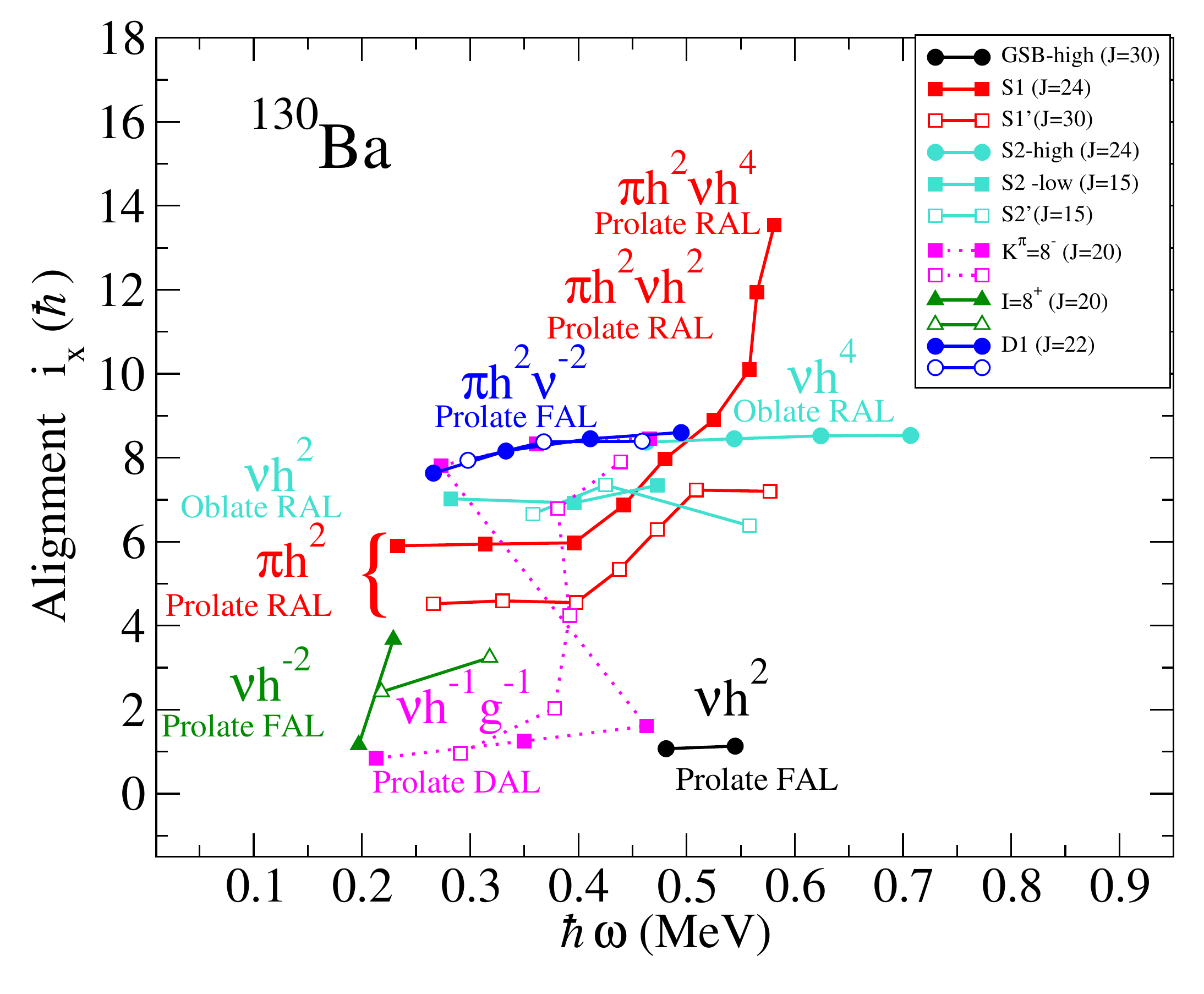}
\includegraphics[angle=0,width=8.5cm]{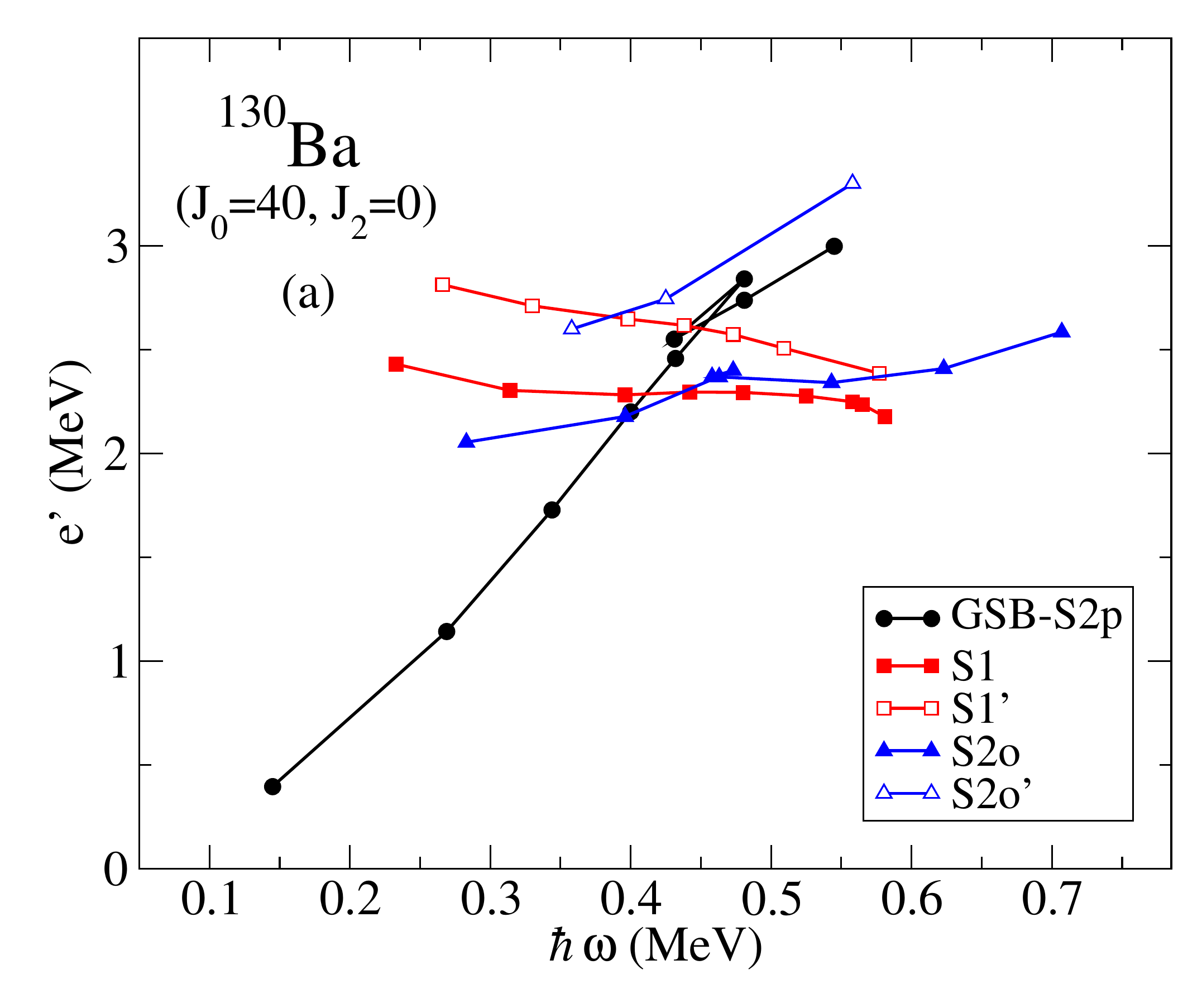}
\caption{(Color online) (a) Experimental total spin $I$ versus rotational frequency $\hbar \omega$  for selected bands of $^{130}$Ba. (b) Experimental quasiparticle alignments for selected bands $^{130}$Ba. (c) Experimental Routhians for selected bands of $^{130}$Ba.}
\label{fig9}
\end{figure*}

\begin{figure}[]
\hskip -6.5cm
\includegraphics[angle=0,width=15.5cm]{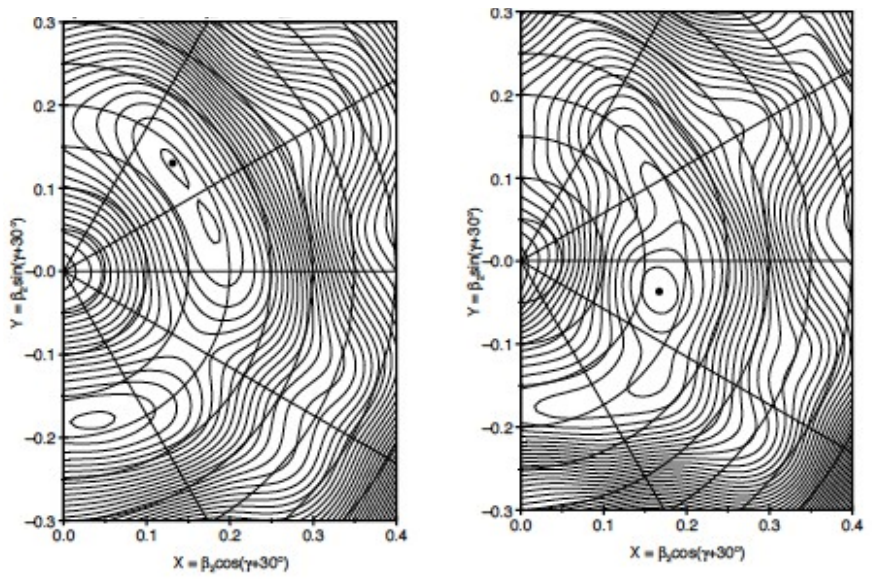}
\vskip-16.cm
\caption{(Color online) TRS calculations for the positive-parity even-spin configurations at rotational frequency 0.05 MeV/$\hbar$ (left) and 0.35 MeV/$\hbar$ (right), showing the shallow prolate and oblate triaxial minima.}
\label{fig10}
\end{figure}

\begin{figure*}[]
\centering
\includegraphics[width=6.5in]{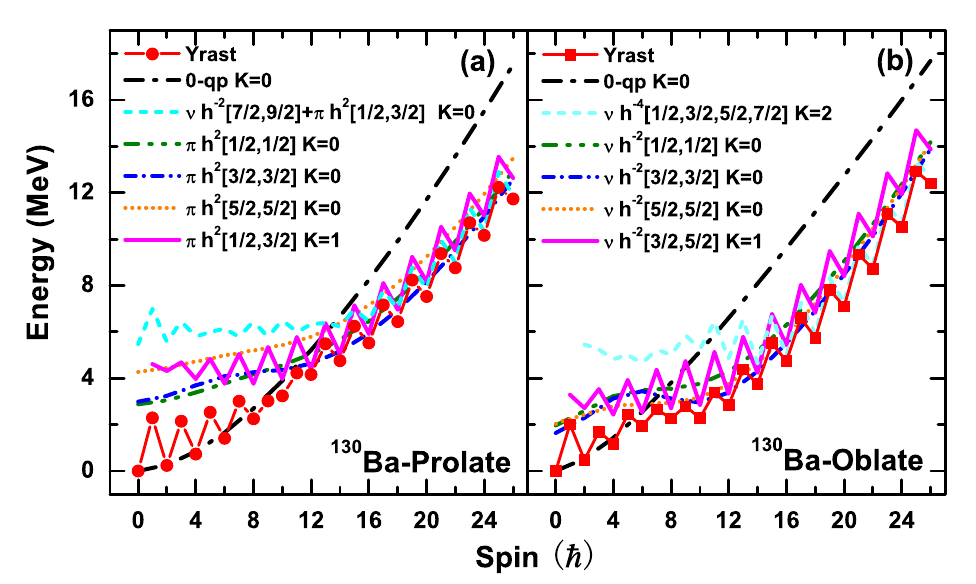}
\caption{ (Color online) Band diagrams for $^{130}$Ba with assuming prolate (a) and oblate shape (b), respectively.}
\label{fig11}
\end{figure*}